\documentclass[11pt]{article}
\usepackage{a4wide}
\usepackage{amsmath,amssymb}
\usepackage{simplewick}
\usepackage{graphicx}
\usepackage{color}
\usepackage{braket}
\usepackage{comment}
\usepackage[bookmarks=true,bookmarksnumbered=true,setpagesize=false]{hyperref}

\DeclareMathOperator{\sgn}{sgn}

\renewcommand{\t}{\text{t}}

\newcommand{\sr}{\stackrel}

\usepackage{enumerate}
\newcommand{\al}[1]{\begin{align}#1\end{align}}
\newcommand{\als}[1]{\begin{align*}#1\end{align*}}
\newcommand{\ov}{\over}
\newcommand{\nn}{\nonumber\\}
\newcommand{\tx}{\text}

\newcommand{\paren}[1]{\left(#1\right)}
\newcommand{\pn}[1]{\left(#1\right)}
\newcommand{\sqbr}[1]{\left[#1\right]}
\newcommand{\ab}[1]{\left|#1\right|}
\newcommand{\abb}[1]{\left\|#1\right\|}
\newcommand{\br}[1]{\left\{#1\right\}}
\newcommand{\fn}[1]{\!\left(#1\right)}

\newcommand{\Paren}[1]{\bigl(#1\bigr)}
\newcommand{\Pn}[1]{\bigl(#1\bigr)}
\newcommand{\Sqbr}[1]{\bigl[#1\bigr]}
\newcommand{\Ab}[1]{\bigl|#1\bigr|}
\newcommand{\Abb}[1]{\bigl\|#1\bigr\|}

\newcommand{\Fn}[1]{\bigl({#1}\bigr)}

\newcommand{\bs}{\boldsymbol}
\newcommand{\bh}[1]{\boldsymbol{\hat#1}}

\newcommand{\df}{\text{d}}

\newcommand{\mc}{\mathcal}

\newcommand{\bmat}[1]{\begin{bmatrix}#1\end{bmatrix}}

\newcommand{\p}{\partial}

\newcommand{\Or}[1]{\mathcal O\!\left(#1\right)}
\newcommand{\h}{\hat}

\newcommand{\pr}{\prime}

\newcommand{\commutator}[2]{\left[#1\,,\,#2\right]}
\newcommand{\Commutator}[2]{\bigl[#1\,,\,#2\bigr]}

\newcommand{\wt}{\widetilde}

\newcommand{\wh}{\widehat}

\newcommand{\dKet}[1]{\left.#1\!\right\rangle}
\newcommand{\dBra}[1]{\left\langle\!#1\right.}
\newcommand{\dBraket}[1]{\left\langle\!#1\!\right\rangle}

\newcommand{\dK}[1]{\left.\Ket{#1}\!\right\rangle}
\newcommand{\dB}[1]{\left\langle\!\Bra{#1}\right.}
\newcommand{\dBK}[1]{\left\langle\!\Braket{#1}\!\right\rangle}
\newcommand{\dBk}[1]{\left\langle\!\Braket{#1}\right.}

\newcommand{\ola}{\overleftarrow}

\newcommand{\nab}{\bs\nabla}
\newcommand{\cA}{\widehat{\mathcal A}}

\newcommand{\tf}{\textsf}
\newcommand{\vep}{\varepsilon}

\usepackage[normalem]{ulem}

\begin{document}

\title{A complete set of Lorentz-invariant wave packets\\
and modified uncertainty relation}

\author{Kin-ya Oda\thanks{E-mail: \tt odakin@lab.twcu.ac.jp} \mbox{}
and Juntaro Wada\thanks{E-mail: \tt wada-juntaro@g.ecc.u-tokyo.ac.jp}}
\maketitle
\begin{center}
\it
$^*$ Department of Mathematics, Tokyo Woman's Christian University, Tokyo 167-8585, Japan\\
$^\dagger$Department of Physics, University of Tokyo, Tokyo 113-0033, Japan\\
\end{center}
\rm
\begin{abstract}\noindent
We define a set of fully Lorentz-invariant wave packets and show that it spans the corresponding one-particle Hilbert subspace, and hence the whole Fock space as well, with a manifestly Lorentz-invariant completeness relation (resolution of identity). The position-momentum uncertainty relation for this Lorentz-invariant wave packet deviates from the ordinary Heisenberg uncertainty principle, and reduces to it in the non-relativistic limit.
\end{abstract}

\newpage
\tableofcontents
\newpage

\section{Introduction}
Wave packets are one of the most fundamental building blocks of quantum field theory. We never observe a plane-wave state of, say, zero and infinite uncertainties of momentum and position, respectively. The plane-wave construction necessarily yields a square of the energy-momentum delta function in the probability, which hence is always divergent and is \emph{more a mnemonic than a derivation} (quoted from Sec.~3.4 in textbook~\cite{Weinberg:1995mt}).

However so far, it has been widely believed that there are no intrinsically new phenomena appearing from a wave-packet construction, but recent developments imply that it might play important roles in vast areas of science; see e.g.\ references in Introduction in Ref.~\cite{Ishikawa:2021bzf}.

Up to now, the wave-packet S-matrix has been computed using a complete basis of Gaussian wave-packets; see e.g.\ Refs.~\cite{Ishikawa:2005zc,Ishikawa:2018koj,Ishikawa:2020hph,Ishikawa:2021bzf}.
The Gaussian basis is constructed from a Gaussian wave packet that evolves in time $t$ as $e^{-i\sqrt{m^2+\bs p^2}t}$ for each plane-wave mode $\bs p$, and is not manifestly Lorentz covariant nor invariant.
To fully exploit the Lorentz covariance of S-matrix in quantum field theory, it is desirable to have a complete basis of \emph{Lorentz-invariant} wave packets. This is what we propose in this paper.

Here we stress a viewpoint that the Gaussian basis is equivalent to a complete set of coherent states in the position-momentum phase space (see Refs.~\cite{Kaiser:1977ys,Kaiser:1978jg,Ali:1985vj,Mostafazadeh:2006gx,Kowalski:2018xsw} and references therein for works on coherent states in the context of the relativistic quantum mechanics\footnote{
It is well known that relativistic quantum mechanics is pathological and that quantum field theory is needed to remedy it; see e.g.\ Refs.~\cite{Newton:1949cq,Wightman:1962sk,Hegerfeldt:1974qu,Hegerfeldt:1980sq,Kalnay} and references therein for related discussions; we clarify our standpoint in Sec.~\ref{Gaussian and coherent section}.
}). Guided by this equivalence, we develop the complete basis of Lorentz-invariant wave packets which is directly applicable in quantum field theory.

Our proposal is also inspired by the ``relativistic Gaussian packet''~\cite{Naumov:2010um,Naumov:2020yyv} developed by Naumov and Naumov (see also Refs.~\cite{AlHashimi:2009bb,Korenblit:2017pto,Kosower:2018adc}), and can be viewed as its generalization to form the complete set: From this viewpoint, our work can be interpreted as a new introduction of a spacetime center of wave packet as an independent variable, which is integrated over a spacelike hyperplane in the completeness relation along with a center of momentum.

It is worth mentioning that our Lorentz-invariant wave packet, when written in momentum space, is essentially the same as the one proposed in Refs.~\cite{Kaiser:1977ys,Kaiser:1978jg}. What is new in this paper in this respect is that we have also defined the wave function in position space and have computed it into an explicit closed form. Thanks to this, we can consider various limits to develop physical intuition. The momentum uncertainty we obtained is in agreement with that in Refs.~\cite{Kaiser:1977ys,Kaiser:1978jg}, whereas the expectation value and uncertainty of the position on a constant time slice are obtained for the first time in this paper.

The organization of this paper is as follows:
In Sec.~\ref{Gaussian and coherent section}, we review the plane-wave basis and the Gaussian basis, as well as the equivalence of the latter to the coherent basis, in order to spell out our notation.
In Sec.~\ref{Lorentz-invariant wave packet section}, we present the Lorentz-invariant wave packet that we propose.
In Sec.~\ref{uncertainty section}, we show the uncertainty relation on this state. In particular, we show that the position-momentum uncertainty deviates from that of the Heisenberg uncertainty principle, while the former reduces to the latter in the non-relativistic limit.
In Sec.~\ref{completeness section}, we prove that these Lorentz invariant wave packets form a complete basis and that the completeness relation can be written in manifestly Lorentz-invariant fashion. As an example, we also show how a scalar field is expanded by this basis of Lorentz-invariant wave packets.
In Appendix~\ref{coherent appendix}, we show some of the known facts on the coherent states.
In Appendix~\ref{basic integral}, we present detailed computations for integrals that we encounter in the main text.

\section{Gaussian basis and coherent states}\label{Gaussian and coherent section}
Here in order to spell out our notation, we review basic known facts about the plane-wave basis and the Gaussian one, as well as the equivalence of the latter to the coherent one in the position-momentum space.
\subsection{Plane-wave basis}
We work in $D=d+1$ dimensional flat spacetime with a metric convention $(-,+,\dots,+)$ such that $e^{ip\cdot x}=e^{-ip^0x^0+i\bs p\cdot\bs x}$ and $p^2=-\paren{p^0}^2+\bs p^2$, where $p_0=-p^0$ and a bold letter denotes a $d$-vector $\bs p=\pn{p^1,\cdots,p^d}=\pn{p_1,\cdots,p_d}$, etc.
Here and hereafter, $x=\pn{x^0,\bs x}$ are coordinates in an arbitrary reference frame.
When $p$ is on-shell, $p^2=-m^2$, $p^0=E_{\bs p}:=\sqrt{\bs p^2+m^2}$, and $e^{ip\cdot x}=e^{-iE_{\bs p}x^0+i\bs p\cdot\bs x}$.
Throughout this paper, we take the number of spatial dimensions $d\geq2$, all the momenta to be on-shell, and all the particles massive $m>0$, unless otherwise stated.
In particular, we use both of $p^0=E_{\bs p}$ (and of $P^0=E_{\bs P}$ appearing below) interchangeably.

In this paper, we focus on a free real scalar field that can be expanded in the Schr\"odinger picture as
\al{
\wh\phi\fn{\bs x}
	&=	\int{\df^d\bs p\ov\paren{2\pi}^{d\ov2}\sqrt{2E_{\bs p}}}\paren{\wh a_{\bs p}e^{i\bs p\cdot\bs x}+\wh a_{\bs p}^\dagger e^{-i\bs p\cdot\bs x}},
}
where
$\wh a_{\bs p}^\dagger$ and $\wh a_{\bs p}$ are the creation and annihilation operators that obey
\al{
\commutator{\wh a_{\bs p}}{\wh a_{\bs p'}^\dagger}
	&=	\delta^d\fn{\bs p-\bs p'}\wh 1,&
\tx{others}
	&=	0,
	\label{creation annhilation operators}
}
where $\wh 1$ is the identity operator on the whole Hilbert space, namely the Fock space $\mc H=\oplus_{n=0}^\infty S\sqbr{L^2\fn{\mathbb R^d}}^{\otimes n}$ with $S$ and $L^2\fn{\mathbb R^d}$ being the symmetrization and the free one-particle momentum space, respectively.
On this space, the free Hamiltonian $\wh H_\tx{free}$ can be expressed as
\al{
\wh H_\tx{free}
	&=	\int\df^d\bs p\,E_{\bs p}\wh a^\dagger_{\bs p}\wh a_{\bs p}
}
up to a constant term. Similarly, the generator of the translation in the free theory is
\al{
\bs{\wh P}_\tx{free}
	&=	\int\df^d\bs p\,\bs p\,\wh a^\dagger_{\bs p}\wh a_{\bs p}.
		\label{momentum operator}
}
In the interaction picture,\footnote{\label{spinor and vector}
Throughout this paper, all the operators other than $\wh\phi\fn{x}$, $\wh\psi\fn{x}$, and $\wh A_\mu\fn{x}$ are time-independent ones in the Schr\"odinger picture, unless otherwise stated.
}
\al{
\wh\phi\fn{x}
	&=	e^{i\wh H_\tx{free}x^0}\wh\phi\fn{\bs x}e^{-i\wh H_\tx{free}x^0}
	=	\int{\df^d\bs p\ov\paren{2\pi}^{d\ov2}\sqrt{2E_{\bs p}}}\paren{\wh a_{\bs p}e^{ip\cdot x}+\wh a_{\bs p}^\dagger e^{-ip\cdot x}},
		\label{phi expanded original}
}
where $p^0=E_{\bs p}=\sqrt{\bs p^2+m^2}$ as always.

We are focusing on the real scalar field in this paper because it is straightforward to generalize it to spinor and vector fields:
We may expand these fields (in the interaction picture) as
\al{
\wh\psi\fn{x}
	&=	\sum_s\int{\df^d\bs p\ov\paren{2\pi}^{d\ov2}\sqrt{2E_{\bs p}}}\paren{\wh a_{\bs p,s}e^{ip\cdot x}u\fn{\bs p,s}+\wh a_{\bs p,s}^{c\dagger} e^{-ip\cdot x}v\fn{\bs p,s}},\\
\wh A_\mu\fn{x}
	&=	\sum_s\int{\df^d\bs p\ov\paren{2\pi}^{d\ov2}\sqrt{2E_{\bs p}}}\paren{\wh a_{\bs p,s}e^{ip\cdot x}\epsilon_\mu\fn{\bs p,s}+\wh a_{\bs p,s}^\dagger e^{-ip\cdot x}\epsilon_\mu^*\fn{\bs p,s}}.
}
and may generalize the expressions below by the replacement $\wh a_{\bs p}\to\wh a_{\bs p,s}$, etc.\footnote{
If one would define the position basis as $\dK{x}=\wh\psi^\dagger\fn{x}\Ket{0}$, etc., the ``Lorentz-invariant wave function'' should read the ``Lorentz-covariant wave function'' accordingly. The position basis for the anti-particle should read $\dK{x,c}=\wh\psi\fn{x}\Ket{0}$ in such a case.
}

Throughout this paper, we concentrate on the free one-particle Hilbert subspace $L^2\fn{\mathbb R^d}$ that is spanned by the free one-particle momentum basis\footnote{\label{basis footnote}
Hereafter, a ``basis'' is used as an abbreviation of a ``basis vector'' or ``basis state'' of a Hilbert space, and denotes a state that can be regarded as an eigenstate of an operator that has proper time evolution in a given picture: Namely, a basis evolves as $\Ket{\phi}$, $e^{i\wh H_\tx{free}x^0}\Ket{\phi}$, and $e^{i\wh Hx^0}\Ket{\phi}$ in the Schr\"odinger, interaction, and Heisenberg pictures, respectively, with $\wh H=\wh H_\tx{free}+\wh H_\tx{int}$. For example, $\Ket{\bs p}$ is a basis in the Schr\"odinger picture. Later, we will call, say, $\dK{\bs x}$ the basis, even though it is not an eigenvector of a Hermitian operator but of a non-Hermitian one~\eqref{chi defined}. See e.g.\ Refs.~\cite{Fujikawa:1994yf,Fujikawa:1995hh} for treatment of non-Hermitian operator.
}
unless otherwise stated:
\al{
\Ket{\bs p}
	&=	\wh a^\dagger_{\bs p}\Ket{0},
}
where the vacuum $\Ket{0}$ is defined by $\wh a_{\bs p}\Ket{0}=0$.
In the subspace $L^2\fn{\mathbb R^d}$, the completeness relation (the resolution of identity) reads
\al{
\int\df^d\bs p\Ket{\bs p}\Bra{\bs p}
	&=	\h 1;
}
here and hereafter, $\h 1$ is the identity operator on $L^2\fn{\mathbb R^d}$.
(Mass dimensions are
$
\Sqbr{\wh\phi}
	=	{d-1\ov2}$ and 
$\sqbr{\wh a_{\bs p}}
	=
\Sqbr{\Ket{\bs p}}
	=	-{d\ov2}$.)

We may take an arbitrary spatial hyperplane~$\Sigma$ as the Fourier transform of the momentum space $\mathbb R^d$. More precisely, $L^2\fn{\mathbb R^d}$ is identified to $L^2\fn{\Sigma}$ through the Fourier transformation.
Here, we take an arbitrary reference frame $x=\pn{x^0,\bs x}$, choose $\Sigma$ to be the ($x^0=0$)-hyperplane $\Sigma_{(0)}$, and define the one-particle position basis $\Ket{\bs x}$ on $\Sigma_{(0)}$ by
\al{
\Braket{\bs x|\bs p}
	&:=	{e^{i\bs p\cdot\bs x}\ov\paren{2\pi}^{d\ov2}}.
		\label{position eigenstates defined}
}
We stress that the Schr\"odinger-picture basis $\Ket{\bs x}$ is already specifying the particular frame~$x$ such that the Minkowski space is foliated by the constant-$x^0$ spacelike hyperplanes and that the free one-particle Hilbert subspace is spanned on a Cauchy surface of a constant-$x^0$ hyperplane, which we have chosen to be $\Sigma_{(0)}$.
We also define, on each constant-$x^0$ hyperplane~$\Sigma_{(x^0)}$ under this foliation, a ``one-particle interaction basis'' $\Ket{x}$ by
\al{
\Braket{x|\bs p}
	&:=	{e^{ip\cdot x}\ov\paren{2\pi}^{d\ov2}}
	=	{e^{-iE_{\bs p}x^0+i\bs p\cdot\bs x}\ov\paren{2\pi}^{d\ov2}}.
	\label{interaction picture basis given}
}
Strictly speaking, $\Ket{x}$ should be regarded as spanning the space $\mc K_{(x^0)}$ of positive-energy solutions to the Klein-Gordon equation at $x^0$, given the initial data $L^2\fn{\Sigma_{(0)}}$ on the Cauchy surface $\Sigma_{(0)}$, whereas one would expect that this is equivalent to $L^2\fn{\Sigma_{(x^0)}}$ due to the time-translational invariance of the theory. Hereafter, we write $L^2\fn{\Sigma_{(x^0)}}$ but a cautious reader may recast it into $\mc K_{(x^0)}$.

We can define the formal momentum operator $\bh p$ on $L^2\fn{\mathbb R^d}$ by
\al{
\bh p\Ket{\bs p}
	&:=	\bs p\Ket{\bs p},
}
and the formal position operator $\bh x$ as the generator of momentum translation on $L^2\fn{\mathbb R^d}$ by\footnote{\label{non-normalizability}
As always, this is written as an operator relation that denotes, for any normalizable physical state $\Ket{\psi}$,
$
\Bra{\bs p}\bh x\Ket{\psi}
	=	i\nab_{\bs p}\Braket{\bs p|\psi}$.
Strictly speaking, the eigenstates of $\bh x$ and $\bh p$ are not an element of $L^2\fn{\Sigma_{(0)}}$ and $L^2\fn{\mathbb R^d}$, respectively, hence the wording ``formal''; see below for more discussion.
}
\al{
\Bra{\bs p}\bh x
	&:=	i\nab_{\bs p}\Bra{\bs p},
	\label{xhat defined}
}
where $\pn{\nab_{\bs p}}_i={\p\ov\p p^i}$.
They satisfy the canonical commutator $\commutator{\h x_i}{\h p_j}=i\delta_{ij}\h 1$, where $\h 1$ has been the identity operator on the one-particle subspace $L^2\fn{\mathbb R^d}$ as said above.
Since we have chosen $\Sigma_{(0)}$ to be the Fourier transform of the momentum space $\mathbb R^d$, we also obtain
\al{
\bh x\Ket{\bs x}
	&=	\bs x\Ket{\bs x}
		\label{x eigensystem}
}
on $L^2\fn{\Sigma_{(0)}}$, which is consistent with Eqs.~\eqref{position eigenstates defined} and \eqref{xhat defined}.
We note that $\bh x$ and $\Ket{\bs x}$ are the time-independent ones in the Schr\"odinger picture by construction; recall footnote~\ref{spinor and vector}.

Here we stress that the position and momentum bases $\Ket{\bs x}$ and $\Ket{\bs p}$ have infinite norms $\Braket{\bs x|\bs x}=\infty$ and $\Braket{\bs p|\bs p}=\infty$, respectively, so that they do not belong to $L^2\fn{\mathbb R^d}$ nor to $L^2\fn{\Sigma_{(0)}}$. We never realize $\Ket{\bs x}$ nor $\Ket{\bs p}$ in any physical experiment. The formal position and momentum operators $\bh x$ and $\bh p$ and their eigenbases $\Ket{\bs x}$ and $\Ket{\bs p}$, respectively, are mere mathematical tools to write down their expectation values basis-independently for any shape of normalizable wave packet $\Ket{\psi}$ in $L^2\fn{\Sigma_{(0)}}$ or $L^2\fn{\mathbb R^d}$ as
\al{
\Braket{\h x_i}_{\Ket{\psi}}
	&=	{\Bra{\psi}\h x_i\Ket{\psi}\ov\Braket{\psi|\psi}}
	=	{1\ov\Braket{\psi|\psi}}\int\df^3\bs x\ab{\psi\fn{\bs x}}^2x_i
	=	{1\ov\Braket{\psi|\psi}}\int\df^3\bs p\,\psi^\dagger\fn{\bs p}\pn{i{\p\ov\p p^i}}\psi\fn{\bs p},\\
\Braket{\h p_i}_{\Ket{\psi}}
	&=	{\Bra{\psi}\h p_i\Ket{\psi}\ov\Braket{\psi|\psi}}
	=	{1\ov\Braket{\psi|\psi}}\int\df^3\bs p\ab{\psi\fn{\bs p}}^2p_i
	=	{1\ov\Braket{\psi|\psi}}\int\df^3\bs x\,\psi^\dagger\fn{\bs x}\pn{-i{\p\ov\p x^i}}\psi\fn{\bs x},
}
where $\psi\fn{\bs x}=\Braket{\bs x|\psi}$, $\psi\fn{\bs p}=\Braket{\bs p|\psi}$, and $\Braket{\psi|\psi}=\int\df^3\bs x\ab{\psi\fn{\bs x}}^2=\int\df^3\bs p\ab{\psi\fn{\bs p}}^2<\infty$.
This fact of the non-normalizability of basis is indeed one of the motivations of the Gaussian construction and its Lorentz-invariant generalization presented in this paper.

The formal operator $\bh x$ is first defined as the generator of momentum translation~\eqref{xhat defined}, and is associated with a particular foliation of spacetime through Eq.~\eqref{x eigensystem} such that $\Sigma_{(0)}$ is chosen as the Fourier transform of the momentum space $\mathbb R^d$ via Eq.~\eqref{position eigenstates defined}. Once this association with the position space is fixed, $\bh x$ is tied to the particular reference frame, with its unit spatial volume $\df^d\bs x$ manifestly violating the Lorentz invariance. The position operator $\bh x$ is \emph{not} covariant by construction; see also Ref.~\cite{Kalnay} for a review on the Lorentz non-covariance from the point of view of relativistic quantum mechanics.

We may regard $\bh p$ as a restriction of the momentum operator~\eqref{momentum operator} to the free one-particle subspace $L^2\fn{\mathbb R^d}$: Schematically,
\al{
\bs{\wh P}_\tx{free}
	&=	
	\bmat{
	0&0&0&\cdots\\
	0&\bh p&0&\cdots\\
	0&0&*&\cdots\\
	\vdots&\vdots&\vdots&\ddots}
\qquad\tx{on}\qquad
	\bmat{\Ket{0}\\
		\tx{1-particle subspace}\\
		*\\
		\vdots}.
}
Similarly, when restricted to the one-particle subspace,
\al{
\h H_\tx{free}
	&:=	\left.\wh H_\tx{free}\right|_{\tx{on }L^2\fn{\mathbb R^d}}=E_{\bh p}=\sqrt{m^2+\bh p^2}.
}
It also follows that, on $L^2\fn{\Sigma_{(0)}}$,
\al{
\Bra{\bs x}\bh p
	&=	-i\nab\Bra{\bs x},&
\bh p\Ket{\bs x}
	&=	i\Ket{\bs x}\overleftarrow\nab,
		\label{momentum basis}
}
such that
$
\Bra{\bs x}\bh p\Ket{\bs p}
	=	-i\nab\Braket{\bs x|\bs p}
	=	-i\nab{e^{i\bs p\cdot\bs x}\ov\paren{2\pi}^{d\ov2}}
	=	\bs p{e^{i\bs p\cdot\bs x}\ov\paren{2\pi}^{d\ov2}}
	=	\bs p\Braket{\bs x|\bs p},
$
with $\nab_i:=\p/\p x^i$.


We may relate the bases of $L^2\fn{\Sigma_{(x^0)}}$ and $L^2\fn{\Sigma_{(0)}}$ by
\al{
\Ket{x}
	&=	e^{iE_{\bh p}x^0}\Ket{\bs x}.
}
The position operator $\bh x$ can be trivially extended to $L^2\fn{\Sigma_{(x^0)}}$:
\al{
\bh x\Ket{x}
	&=	\bh xe^{iE_{\bh p}x^0}\Ket{\bs x}
	=	\paren{\bs x-{\bh p\ov E_{\bh p}}x^0}\Ket{x},
		\label{eigenvalue of interaction-picture x state}
}
where we used $\commutator{\h x_i}{f\fn{\bh p}}=i{\p f\ov\p p^i}\fn{\bh p}$.

On $L^2\fn{\Sigma_{(0)}}$, the plane-wave normalization is\footnote{
In Refs.~\cite{Hegerfeldt:1974qu,Hegerfeldt:1980sq} it has been shown that (what-we-call) ``strict localization'', which requires a wave function $\psi\fn{\bs x}=\Braket{\bs x|\psi}$ in $L^2\fn{\Sigma_{(0)}}$ to vanish everywhere outside a finite region $V\subset\Sigma_{(0)}$, cannot be consistent with what the authors call ``causality'', which we will refer to the ``Hegerfeldt causality''. The Hegerfeldt causality holds when the following is satisfied: If $\psi$ is strictly localized to $V$, then there should exist $r$ that makes $\Bra{\bs x}e^{-i\h H_\tx{free}x^0+i\bh p\cdot\bs a}\Ket{\psi}=0$ for all $\bs x\in V$ for all $\bs a$ with $\bs a>r$ at any later time $x^0>0$. The authors have proven that the Hegerfeldt causality is necessarily violated~\cite{Hegerfeldt:1974qu,Hegerfeldt:1980sq}. From Eq.~\eqref{delta function inner product}, the position basis $\Ket{\bs x'}$, interpreted as a wave function of $\bs x$ in $L^2\fn{\Sigma_{(0)}}$ (closing our eyes on the fact that it cannot the case due to its non-normalizability), is strictly localized. Therefore, it obeys the proven violation of the Hegerfeldt causality. On the other hand, both the Gaussian wave packet~\eqref{Gaussian wave function} and our extension~\eqref{Lorentz-invariant wave function} have exponentially small but non-zero tail outside any finite region $V$ from the beginning on $\Sigma_{(0)}$ and later. Therefore, they evade the condition of strict localization of the Hegerfeldt causality from the beginning.
}
\al{
\Braket{\bs x|\bs x'}
	&=	\delta^d\fn{\bs x-\bs x'}.
		\label{delta function inner product}
}
We may also formally write down the inner product of bases of $L^2\fn{\Sigma_{(x^0)}}$ and of  $L^2\fn{\Sigma_{(x^{\pr0})}}$,
\al{
\Braket{x|x'}
	&=	\int\df^d\bs p\Braket{x|\bs p}\Braket{\bs p|x'}
	=	\int{\df^d\bs p\ov\paren{2\pi}^d}e^{ip\cdot\paren{x-x'}}
	=	\int{\df^d\bs p\ov\paren{2\pi}^d}e^{-iE_{\bs p}\paren{x^0-x^{\pr0}}+i\bs p\cdot\paren{\bs x-\bs x'}}.
}
The completeness relations on $L^2\fn{\Sigma_{(0)}}$ and on $L^2\fn{\Sigma_{(x^0)}}$ are, respectively,
\al{
\int\df^d\bs x\Ket{\bs x}\Bra{\bs x}
	&=	\h1,&
\int\df^d\bs x\Ket{x}\Bra{x}
	&=	\h1.
}
Mass dimensions are
$
\Sqbr{\Ket{\bs x}}
	=	\Sqbr{\Ket{x}}
	=	{d\ov2}$.

\subsection{Lorentz-friendly bases}
From here we start to deviate from the standard notation in the literature. What-we-will-call ``Lorentz-friendly basis'' is essentially the same as the basis proposed by Newton and Wigner~\cite{Newton:1949cq}, which is later complemented in terms of the Euclidean group by Wightman~\cite{Wightman:1962sk}.

We define a ``Lorentz-friendly'' annihilation operator on $\mc H$:
\al{
\wh\alpha_{\bs p}
	&=	\sqrt{2E_{\bs p}}\,\wh a_{\bs p},
		\label{Lorentz-friendly annihilation operator}
}
which gives
\al{
\commutator{\wh\alpha_{\bs p}}{\wh\alpha_{\bs p'}^\dagger}
	&=	2E_{\bs p}\,\delta^d\fn{\bs p-\bs p'}\wh 1,&
\tx{others}
	&=	0,
}
and
\al{
\wh\phi\fn{x}
	&=	\int{\df^d\bs p\ov\paren{2\pi}^{d\ov2}2E_{\bs p}}\paren{\wh\alpha_{\bs p}e^{ip\cdot x}+\wh\alpha_{\bs p}^\dagger e^{-ip\cdot x}}\nn
	&=	\int{\df^Dp\ov\paren{2\pi}^{d\ov2}}\delta\fn{p^2+m^2}\theta\fn{p^0}\paren{\wh\alpha_{\bs p}e^{ip\cdot x}+\wh\alpha_{\bs p}^\dagger e^{-ip\cdot x}},
		\label{phi expanded}
}
where the Lorentz invariance is made manifest in the last expression by letting $p$ off-shell.

We define a Lorentz-friendly momentum basis that spans $L^2\fn{\mathbb R^d}$:
\al{
\dKet{\Ket{\bs p}}
	&:=	\wh\alpha_{\bs p}\Ket{0}
	=	\sqrt{2E_{\bs p}}\Ket{\bs p}
}
such that
\al{
\dBraket{\Braket{\bs p|\bs p'}}
	&=	2E_{\bs p}\,\delta^d\fn{\bs p-\bs p'},&
\int{\df^d\bs p\ov 2E_{\bs p}}\dKet{\Ket{\bs p}}\dBra{\Bra{\bs p}}
	&=	\h 1.
		\label{completeness of L-friendly p-basis}
}
Mass dimensions are
$
\sqbr{\alpha_{\bs p}}
	=	\Sqbr{\dKet{\Ket{\bs p}}}
	=	-{d-1\ov2}$.
This completeness is the same as Eq.~(1) in Ref.~\cite{Newton:1949cq} up to the factor 2 (which will not be mentioned hereafter).

We also define Lorentz-friendly position bases in $L^2\fn{\Sigma_{(0)}}$ and in $L^2\fn{\Sigma_{(x^0)}}$, respectively:\footnote{
See footnote~\ref{basis footnote}.
}
\al{
\dKet{\Ket{\bs x}}
	&:=	\wh\phi\fn{\bs x}\Ket{0}
	=	\int{\df^d\bs p\ov\paren{2\pi}^{d\ov2}2E_{\bs p}}
			 e^{-i\bs p\cdot\bs x}\dKet{\Ket{\bs p}},\\
\dKet{\Ket{x}}
	&:=	\wh\phi\fn{x}\Ket{0}
	=	\int{\df^d\bs p\ov\paren{2\pi}^{d\ov2}2E_{\bs p}}
			 e^{-ip\cdot x}\dKet{\Ket{\bs p}}.
			 \label{LF x basis}
}
Mass dimensions are
$
\Sqbr{\dKet{\Ket{\bs x}}}
	=	\Sqbr{\dKet{\Ket{x}}}
	=	{d-1\ov2}$.
Here, $\dKet{\Ket{\bs x}}$ and $\dKet{\Ket{x}}$ are generalizations of the one-particle position bases in the Schr\"odinger and interaction pictures, respectively.
They satisfy
\al{
\dBraket{\Braket{\bs x|\bs p}}
	&=	{e^{i\bs p\cdot\bs x}\ov\paren{2\pi}^{d\ov2}}
	=	\Braket{\bs x|\bs p},&
\dBraket{\Braket{x|\bs p}}
	&=	{e^{ip\cdot x}\ov\paren{2\pi}^{d\ov2}}
	=	\Braket{x|\bs p}.
}
Now we can write the field operator on $\mc H$ as
\al{
\wh\phi\fn{x}
	&=	\int{\df^d\bs p\ov2E_{\bs p}}\paren{
			\dBraket{\Braket{x|\bs p}}\wh\alpha_{\bs p}
			+\wh\alpha_{\bs p}^\dagger\dBraket{\Braket{\bs p|x}}
			}.
			\label{expansion by alpha}
}
Note that a wave function $\dBk{x|\Phi}$ is equivalent to the one given in Eq.~(2) in Ref.~\cite{Newton:1949cq}.

On $L^2\fn{\Sigma_{(0)}}$, we may formally write
\al{
\dBra{\Bra{\bs x}}
	&=	\Bra{\bs x}{1\ov\sqrt{2E_{\bh p}}}
	=	{1\ov\sqrt{2\sqrt{-\nab^2+m^2}}}\Bra{\bs x}.
}
The normalization on $L^2\fn{\Sigma_{(0)}}$ is
\al{
\dBK{\bs x|\bs x'}
	&=	\int{\df^d\bs p\ov2E_{\bs p}}\dBraket{\Braket{\bs x|\bs p}}\dBraket{\Braket{\bs p|\bs x'}}
	=	\int{\df^d\bs p\ov\paren{2\pi}^d2E_{\bs p}}e^{i\bs p\cdot\paren{\bs x-\bs x'}},
		\label{double braket inner product}
}
and we may again write down the inner product of bases of $L^2\fn{\Sigma_{(x^0)}}$ and of $L^2\fn{\Sigma_{(x^{\pr0})}}$:\footnote{
From this, the Feynman propagator is given by
\als{
D_\tx{F}\fn{x-x'}
	&=	\theta\fn{x^0-x^{\pr0}}\dBK{x|x'}+\theta\fn{x^{\pr0}-x^0}\dBK{x'|x}
	=	\int{\df^Dp\ov\paren{2\pi}^D}e^{i p\cdot\pn{x-x'}}
		{-i\ov p^2+m^2-i\epsilon}.
}
One may find its explicit form as a function of $\pn{x-x'}^2$ e.g.\ in Ref.~\cite{Zhang:2008jy}.
}
\al{
\dBK{x|x'}
	&=	\int{\df^d\bs p\ov2E_{\bs p}}\dBraket{\Braket{x|\bs p}}\dBraket{\Braket{\bs p|x'}}
	=	\int{\df^d\bs p\ov\paren{2\pi}^d2E_{\bs p}}e^{ip\cdot\paren{x-x'}}\nn
	&=	\int{\df^Dp\ov\paren{2\pi}^d}\theta\fn{p^0}\delta\fn{p^2+m^2}e^{ip\cdot\paren{x-x'}}.
}

On $L^2\fn{\Sigma_{(0)}}$, the completeness relation becomes
\al{
\h	1
	&=	\int\df^d\bs x\sqrt{2E_{\bh p}}\dKet{\Ket{\bs x}}\dBra{\Bra{\bs x}}\sqrt{2E_{\bh p}}&
	&=	\int\df^d\bs x\sqbr{
				\dKet{\Ket{\bs x}}\sqrt{2\sqrt{m^2-\overleftarrow\nab^2}}
				}
			\sqbr{
				\sqrt{2\sqrt{m^2-\nab^2}}\dBra{\Bra{\bs x}}
				}\nn
	&=	\int\df^d\bs x\,2E_{\bh p}\dKet{\Ket{\bs x}}\dBra{\Bra{\bs x}}&
	&=	\int\df^d\bs x\sqbr{
			\dKet{\Ket{\bs x}}2\sqrt{m^2-\overleftarrow\nab^2}
			}
			\dBra{\Bra{\bs x}}\nn
	&=	\int\df^d\bs x\dKet{\Ket{\bs x}}\dBra{\Bra{\bs x}}2E_{\bh p}&
	&=	\int\df^d\bs x
			\dKet{\Ket{\bs x}}
			\sqbr{2\sqrt{m^2-\nab^2}\dBra{\Bra{\bs x}}},
			\label{completeness in Lorentz-friendly x}
}
which can be checked by sandwiching both-hand sides by $\dBra{\Bra{\bs p}}$ and $\dKet{\Ket{\bs p'}}$.
The same relation holds on $L^2\fn{\Sigma_{(x^0)}}$ when we replace $\dKet{\Ket{\bs x}}$ and $\dBra{\Bra{\bs x}}$ by $\dKet{\Ket{x}}$ and $\dBra{\Bra{x}}$, respectively, because the factors $e^{\pm i\sqrt{m^2+\nab^2}x^0}$ cancel out each other.
Now we may rewrite the above completeness relation in a manifestly Lorentz-invariant fashion\footnote{
More precisely, the relation becomes Lorentz invariant when both-hand sides are sandwiched by the basis states $\dB{\bs p}$ and $\dK{\bs q}$.
}
on an arbitrary $L^2\fn{\Sigma}$ space, with the same precaution as given after Eq.~\eqref{interaction picture basis given}:
\al{
\h	1
	&=	\int_\Sigma\df^d\Sigma^\mu\dK{x}\sqbr{2i{\p\ov\p x^\mu}\dB{x}}
	=	\int_\Sigma\df^d\Sigma^\mu\sqbr{\dK{x}\pn{-2i\ola{\frac{\p}{\p x^\mu}}}}\dB{x},
		\label{completeness, Lorentz invariant}
}
where $\df^d\Sigma^\mu$ is the surface element normal to $\Sigma$.
(In the language of differential forms, it is nothing but the induced volume element $\df^d\Sigma^\mu=-\,{\star\df x^\mu}$, with $\star$ denoting the Hodge dual; in the flat spacetime, we get $\star\df x_\mu={1\ov d!}\epsilon_{\mu\mu_1\dots\mu_d}\df x^{\mu_1}\wedge\cdots\wedge\df x^{\mu_d}$ with $\epsilon_{01\dots d}=1$.)

Physically, a probability density $P\fn{x}$ (per unit volume $\df^d\bs x$) of observing the particle at a position $\bs x$ at time $x^0$ for a (normalized) wave packet $\Ket{\psi}$ is given by the expectation value of the projector $\Ket{x}\Bra{x}$ on $L^2\fn{\Sigma_{(x^0)}}$:\footnote{
We are working in the interaction picture and hence the time dependence of the wave function is $e^{i\wh H_\tx{free}x^0}e^{-i\wh Hx^0}\Ket{\psi}$ with $\wh H=\wh H_\tx{free}+\wh H_\tx{int}$. Throughout this paper, we are neglecting interactions, $\wh H_\tx{int}=0$, and hence it suffices to treat the time dependence as in the main text.
}
\al{
P\fn{x}
	&=	\Braket{\psi|x}\Braket{x|\psi}
	=	\Bra{\psi}\sqrt{2E_{\bh p}}\dK{x}\dB{x}\sqrt{2E_{\bh p}}\Ket{\psi}
	=	\ab{\sqrt{2\sqrt{m^2-\nab^2}}\dBk{x|\psi}}^2.
	\label{probability density}
}
Note that the probability density is not mere an absolute-square of the wave function.

The following relations on $L^2\fn{\mathbb R^d}$ may be useful:
\al{
\dBra{\Bra{\bs p}}\bh x
	&
	=	\sqrt{2E_{\bs p}}\,i\nab_{\bs p}\Bra{\bs p}
	=	i\nab_{\bs p}\dBra{\Bra{\bs p}}
		-i{\bs p\ov2E_{\bs p}^2}\dBra{\Bra{\bs p}},
		\label{position on p-space}
}
and we get
\al{
\bh x\dKet{\Ket{\bs x}}
	&=	\paren{\bs x-i{\bh p\ov2E_{\bh p}^2}}\dKet{\Ket{\bs x}},
		\label{eigenvalue of non-Hermitian x}\\
\bh x\dKet{\Ket{x}}
	&=	\paren{\bs x-{\bh p\ov E_{\bh p}}x^0-i{\bh p\ov2E_{\bh p}^2}}\dKet{\Ket{x}},
}
on $L^2\fn{\Sigma_{(0)}}$ and on $L^2\fn{\Sigma_{(x^0)}}$, respectively.
The relation~\eqref{position on p-space} is equivalent to Eq.~(11) in Ref.~\cite{Newton:1949cq} up to the metric sign convention.
 
We may formally define \emph{non-}Hermitian position-like operator:\footnote{
This operator has been discussed in Ref.~\cite{Korenblit:2017pto} and references therein, where $\bh\chi$ is treated as self-adjoint.
Our claim differs in that $\bh\chi$ is manifestly non-self-adjoint.
}
\al{
\bh\chi
	&:=	\sqrt{2E_{\bh p}}\bh x{1\ov\sqrt{2E_{\bh p}}}
	=	\bh x-i{\bh p\ov2E_{\bh p}^2},
		\label{chi defined}
}
which satisfies
\al{
\dB{\bs x}\bh\chi
	&=	\bs x\dB{\bs x},&
\bh\chi^\dagger\dK{\bs x}
	&=	\bs x\dK{\bs x},\label{chi operator}\\
\dB{\bs p}\bh\chi^\dagger
	&=	i\nab_{\bs p}\dB{\bs p},&
\bh\chi\dK{\bs p}
	&=	\dK{\bs p}\paren{-i\ola\nab_{\bs p}},
}
on $L^2\fn{\Sigma_{(0)}}$ and on $L^2\fn{\mathbb R^d}$, respectively.
Note that e.g.\ $\dK{\bs x}$ is not an eigenbasis of $\bh\chi$ but of $\bh\chi^\dagger$.
In Eq.~\eqref{chi operator}, the eigenvalues happen to be real for the non-Hermitian operators $\bh\chi$ and $\bh\chi^\dagger$, respectively.

We summarize the results for various bases in Table~\ref{table for x and p}.

\begin{table}\centering
\caption{Eigenvalues of $\bh p$ and $\bh x$ or $\bh\chi$ on various states.
\label{table for x and p}}
\als{
&\tx{State}&	\bh p&	&	\bh\chi=\bh x&-i{\bh p\ov2E_{\bh p}^2}	\\
\hline\hline
\Bra{\bs p}&&
	\Bra{\bs p}\bh p&=\bs p\Bra{\bs p}&
	\Bra{\bs p}\bh x&=i\nab_{\bs p}\Bra{\bs p}\\
\Bra{\bs x}&&
	\Bra{\bs x}\bh p&=-i\nab\Bra{\bs x}&
	\Bra{\bs x}\bh x&=\bs x\Bra{\bs x}\\
\Bra{x}&=\Bra{\bs x}e^{-iE_{\bh p}x^0}&
	\Bra{x}\bh p&=-i\nab\Bra{x}&
	\Bra{x}\paren{\bh x+{\bh p\ov E_{\bh p}}x^0}&=\bs x\Bra{x}\\
\hline
\dBra{\Bra{\bs p}}&=\sqrt{2E_{\bs p}}\Bra{\bs p}&
	\dBra{\Bra{\bs p}}\bh p&=\bs p\dBra{\Bra{\bs p}}&
	\dBra{\Bra{\bs p}}\bh\chi^\dagger&=i\nab_{\bs p}\dBra{\Bra{\bs p}}\\
\dBra{\Bra{\bs x}}&=\Bra{\bs x}{1\ov\sqrt{2E_{\bh p}}}&
	\dBra{\Bra{\bs x}}\bh p&=-i\nab\dBra{\Bra{\bs x}}&
	\dBra{\Bra{\bs x}}\bh\chi&=\bs x\dBra{\Bra{\bs x}}\\
\dBra{\Bra{x}}&=\Bra{\bs x}{e^{-iE_{\bh p}x^0}\ov\sqrt{2E_{\bh p}}}&
	\dBra{\Bra{x}}\bh p&=-i\nab\dBra{\Bra{x}}&
	\dBra{\Bra{x}}\paren{\bh\chi+{\bh p\ov E_{\bh p}}x^0}&=\bs x\dBra{\Bra{x}}\\
\hline\hline
\Ket{\bs p}&&
	\bh p\Ket{\bs p}&=\bs p\Ket{\bs p}&
	\bh x\Ket{\bs p}&=\Ket{\bs p}\paren{-i\ola\nab_{\bs p}}\\
\Ket{\bs x}&&
	\bh p\Ket{\bs x}&=\Ket{\bs x}\paren{i\ola\nab}&
	\bh x\Ket{\bs x}&=\bs x\Ket{\bs x}\\
\Ket{x}&=e^{iE_{\bh p}x^0}\Ket{\bs x}&
	\bh p\Ket{x}&=\Ket{x}\paren{i\ola\nab}&
	\paren{\bh x+{\bh p\ov E_{\bh p}}x^0}\Ket{x}&=\bs x\Ket{x}\\
\hline
\dK{\bs p}&=\sqrt{2E_{\bs p}}\Ket{\bs p}&
	\bh p\dK{\bs p}&=\bs p\dK{\bs p}&
	\bh\chi\dK{\bs p}&=\dK{\bs p}\paren{-i\ola\nab_{\bs p}}\\
\dK{\bs x}&={1\ov\sqrt{2E_{\bh p}}}\Ket{\bs x}&
	\bh p\dK{\bs x}&=\dK{\bs x}\paren{i\ola\nab}&
	\bh\chi^\dagger\dK{\bs x}&=\bs x\dK{\bs x}\\
\dK{x}&={e^{iE_{\bh p}x^0}\ov\sqrt{2E_{\bh p}}}\dK{\bs x}&
	\bh p\dK{x}&=\dK{x}\paren{i\ola\nab}&
	\paren{\bh\chi^\dagger+{\bh p\ov E_{\bh p}}x^0}\dK{x}&=\bs x\dK{x}\\
\hline
}
\flushleft In each set of three rows separated by the horizontal lines, the first, second, and third rows are given in the Hilbert spaces $L^2\fn{\mathbb R^d}$, $L^2\fn{\Sigma_{(0)}}$, and $L^2\fn{\Sigma_{(x^0)}}$, respectively.
\end{table}

\newpage

Finally, we comment on the Lorentz-transformation property of the one-particle momentum and position operators on $L^2\fn{\mathbb R^d}$ or $L^2\fn{\Sigma_{(x^0)}}$.
The Poincar\'e transformation\footnote{
Here we only consider orthochronous $\Lambda$ so that $\wh U_\tx{free}\pn{\Lambda,b}$ is linear and unitary.
}
on the annihilation operator reads (see e.g.\ Ref.~\cite{Weinberg:1995mt})
\al{
\wh a_{\bs p}
	&\to	\wh U_\tx{free}\fn{\Lambda,b}\wh a_{\bs p}\wh U_\tx{free}^\dagger\fn{\Lambda,b}
	=	e^{i\pn{\Lambda p}\cdot b}\sqrt{\pn{\Lambda p}^0\ov p^0}\wh a_{\bs p_\Lambda},\\
\wh\alpha_{\bs p}
	&\to	\wh U_\tx{free}\fn{\Lambda,b}\wh\alpha_{\bs p}\wh U_\tx{free}^\dagger\fn{\Lambda,b}
	=	e^{i\pn{\Lambda p}\cdot b}\wh\alpha_{\bs p_\Lambda},
		\label{LT on alpha}
}
where $\bs p_\Lambda$ denotes the spatial component of $\Lambda p$, namely $\pn{\bs p_\Lambda}^i=\pn{\Lambda p}^i$.
In particular,
\al{
\dK{\bs p}
	&\to	\wh U_\tx{free}\fn{\Lambda,b}\dK{\bs p}
	=		e^{-i\pn{\Lambda p}\cdot b}\dK{\bs p_\Lambda}.
}
We may reinterpret this transformation on states as that on operators
\al{
\h p^i
	&\to	\h p^i_\Lambda
	:=		\wh U_\tx{free}^\dagger\fn{\Lambda,b}\h p^i\wh U_\tx{free}\fn{\Lambda,b},
}
which yields
\al{
\h p^i_\Lambda\dK{\bs p}
	&=	\pn{\Lambda p}^i\dK{\bs p}.
}
The momentum operator is covariant on $L^2\fn{\mathbb R^d}$ in this sense.

The transformation~\eqref{LT on alpha} yields
\al{
\wh\phi\fn{x}
	&\to	\wh U_\tx{free}\fn{\Lambda,b}\wh\phi\fn{x}\wh U_\tx{free}^\dagger\fn{\Lambda,b}
	=	\wh\phi\fn{\Lambda x+b}.
	\label{transformation on fields}
}
From Eqs.~\eqref{LF x basis} and \eqref{transformation on fields}, we see that
\al{
\dK{x}
	&\to
		\wh U_\tx{free}\fn{\Lambda,b}\dK{x}=\dK{\Lambda x+b},
}
Again, reinterpreting this as transformation on operators
\al{
\h\chi^{i\dagger}
	&\to	\h\chi_{\Lambda,b}^{i\dagger}
	:=		\wh U_\tx{free}^\dagger\fn{\Lambda,b}\h\chi^{i\dagger}\wh U_\tx{free}\fn{\Lambda,b},
}
we can show that
\al{
\h\chi_{\Lambda,b}^{i\dagger}\dK{x}
	&=	\pn{\Lambda x+b}^i\dK{x}.
}
We see that, on the Lorentz-friendly basis $\dK{x}$ in $L^2\fn{\Sigma_{(x^0)}}$, the non-Hermitian operator $\bh\chi^\dagger$ is the spatial component of a Poincar\'e-covariant vector.
On the other hand, we clearly see that the physical position operator $\bh x=\bh\chi^\dagger-i{\bh p\ov2E_{\bh p}^2}$ is not a spatial component of a Lorentz-covariant vector due to the second term.
From the view of modern quantum field theory, it is not compulsory that $\bh x$, associated to a particular spacetime foliation $\Sigma_{(x^0)}$ with time slices of constant $x^0$, be a covariant operator. Of course, the whole theory is Lorentz invariant in the sense that the S-matrix, constructed from the covariant quantum fields~\eqref{transformation on fields} defined in the whole space $\mc H$, is Lorentz invariant.

\subsection{Gaussian basis}\label{Gaussian basis section}
We define the Gaussian basis states through a \emph{normalizable}, hence physical, wave function on $L^2\fn{\mathbb R^d}$~\cite{Ishikawa:2005zc}:
\al{
\Braket{\sigma;\bs X,\bs P|\bs p}
	&:=	\paren{\sigma\ov\pi}^{d\ov4}e^{i\bs p\cdot\bs X}e^{-{\sigma\ov2}\paren{\bs p-\bs P}^2},
		\label{Gaussian basis NR}\\
\Braket{\sigma;X,\bs P|\bs p}
	&:=	\paren{\sigma\ov\pi}^{d\ov4}e^{ip\cdot X}e^{-{\sigma\ov2}\paren{\bs p-\bs P}^2},
		\label{Gaussian basis}
}
where $\bs X$ and $\bs P$ are the centers of position and momentum of the wave packet, respectively, at time $x^0=0$ for Eq.~\eqref{Gaussian basis NR} and $x^0=X^0$ for \eqref{Gaussian basis}, while $\sigma>0$ is its width-squared.
We see that these states on $L^2\fn{\mathbb R^d}$ are related by
\al{
\Ket{\sigma;X,\bs P}
	&=	e^{iE_{\bh p}X^0}\Ket{\sigma;\bs X,\bs P}.
}
Due to this dependence on $X^0$, one might want to regard the physical states $\Ket{\sigma;\bs X,\bs P}$ and $\Ket{\sigma;X,\bs P}$ as some bases in the Schr\"odinger and interaction pictures on $L^2\fn{\Sigma_{(0)}}$ and  $L^2\fn{\Sigma_{(X^0)}}$, respectively, through the Fourier transformation.\footnote{
See also footnotes~\ref{basis footnote} and \ref{non-normalizability}.
}
However, when we consider the wave function~\eqref{Gaussian wave function expanded} below, $X^0$ is rather a parameter that specifies the shape of the wave packet, and the time coordinate is $x^0$.

Again through the Fourier transformation, we may map the momentum-space wave functions onto $L^2\fn{\Sigma_{(0)}}$ and $L^2\fn{\Sigma_{(x^0)}}$:
\al{
\Braket{\bs x|\sigma;\bs X,\bs P}
	&=	\int\df^d\bs p\Braket{\bs x|\bs p}\Braket{\bs p|\sigma;\bs X,\bs P}
	=	{1\ov\paren{\pi\sigma}^{d\ov4}}
		e^{i\bs P\cdot\paren{\bs x-\bs X}}e^{-{1\ov2\sigma}\paren{\bs x-\bs X}^2},
		\label{Gaussian wave function}\\
\Braket{x|\sigma;X,\bs P}
	&=	\int\df^d\bs p\Braket{x|\bs p}\Braket{\bs p|\sigma;X,\bs P}
	=	{1\ov\paren{2\pi}^{d\ov2}}\paren{\sigma\ov\pi}^{d\ov4}
		\int\df^d\bs p\,e^{ip\cdot\pn{x-X}}e^{-{\sigma\ov2}\paren{\bs p-\bs P}^2}.
		\label{Gaussian wave function expanded}
}
At the leading saddle-point approximation for large $\sigma$, Eq.~\eqref{Gaussian wave function expanded} reduces to a closed form~\cite{Ishikawa:2018koj}:
\al{
\Braket{x|\sigma;X,\bs P}
	&\to	\pn{1\ov\pi\sigma}^{d/4}
			{1\ov\sqrt{2P^0}}e^{iP\cdot\pn{x-X}}e^{-{\sigma\ov2}\pn{\bs x-\bs X-\bs V\pn{x^0-X^0}}^2},
			\label{closed form of Gaussian wave function}
}
where $P$ is on-shell $P^0=E_{\bs P}$ as always, and we define $\bs V:=\bs P/P^0$.
We see that the center of wave packet moves as $\bs X+\bs V\pn{x^0-X^0}$ when we vary time $x^0$, namely, when we change the time-slice $\Sigma_{(x^0)}$.

The inner product in $L^2\fn{\mathbb R^d}$ is not orthogonal:
\al{
\Braket{\sigma;\bs X,\bs P|\sigma';\bs X',\bs P'}
	&=	\paren{\sigma_\tx{I}\ov\sigma_\tx{A}}^{d\ov4}
		e^{-{1\ov4\sigma_\tx{A}}\paren{\bs X-\bs X'}^2}
		e^{-{\sigma_\tx{I}\ov4}\paren{\bs P-\bs P'}^2}
		e^{{i\ov2\sigma_\tx{I}}\paren{\sigma\bs P+\sigma'\bs P'}\cdot\paren{\bs X-\bs X'}},
}
where $\sigma_\tx{A}:={\sigma+\sigma'\ov2}$ and $\sigma_\tx{I}:=\paren{\sigma^{-1}+\sigma^{\pr-1}\ov2}^{-1}={2\sigma\sigma'\ov\sigma+\sigma'}={\sigma\sigma'\ov\sigma_\tx{A}}$ are the average and the inverse of average of inverse, respectively~\cite{Ishikawa:2020hph}.

It is important that the Gaussian basis, with any fixed $\sigma$, form an (over)complete set in the free one-particle subspace $L^2\fn{\mathbb R^d}$~\cite{Ishikawa:2005zc,Ishikawa:2018koj}:
\al{
\int{\df^d\bs X\,\df^d\bs P\ov\paren{2\pi}^d}\Ket{\sigma;\bs X,\bs P}\Bra{\sigma;\bs X,\bs P}
	&=	\h 1.
	\label{completeness of Gaussian basis}
}
Because any fixed $\sigma$ suffices to provide the complete set spanning $L^2\fn{\mathbb R^d}$, we omit the label $\sigma$ unless otherwise stated hereafter. 
We may expand any wave function (or field configuration obeying the Klein-Gordon equation) $\psi\fn{x}=\Braket{x|\psi}$ by the Gaussian complete set $\Set{\Ket{\bs X,\bs P}}_{\bs X,\bs P}$ that spans $L^2\fn{\mathbb R^d}$:
\al{
\psi\fn{x}
	&=	\int{\df^d\bs X\,\df^d\bs P\ov\paren{2\pi}^d}\Braket{x|\bs X,\bs P}\Braket{\bs X,\bs P|\psi},
}
and conversely, the expansion coefficient may be computed by
\al{
\Braket{\bs X,\bs P|\psi}
	&=	\int\df^d\bs x
		\Braket{\bs X,\bs P|\bs x}
		\Braket{\bs x|\psi}
	=	{1\ov\paren{\pi\sigma}^{d\ov4}}\int\df^d\bs x\,
		e^{-i\bs P\cdot\paren{\bs x-\bs X}}e^{-{1\ov2\sigma}\paren{\bs x-\bs X}^2}
		\Braket{\bs x|\psi}\nn
	&=	\int\df^d\bs p
		\Braket{\bs X,\bs P|\bs p}
		\Braket{\bs p|\psi}
	=	\paren{\sigma\ov\pi}^{d\ov4}
		\int\df^d\bs p\,
		e^{i\bs p\cdot\bs X}e^{-{\sigma\ov2}\paren{\bs p-\bs P}^2}
		\Braket{\bs p|\psi}.
}

\subsection{Coherent states}\label{coherent section}
Here we see that a Gaussian wave packet is indeed a coherent state~\cite{Sudarshan1963,Glauber1963} in the free-one-particle subspace $L^2\fn{\mathbb R^d}$, or equivalently in $L^2\fn{\Sigma_{(0)}}$.

We define an ``annihilation'' operator for a $d$-dimensional harmonic oscillator:
\al{
\bh a
	&:=	\lambda\paren{{\bh x\ov\sqrt{\sigma}}+i\sqrt{\sigma}\bh p},&
\bh a^\dagger
	&=	\lambda^*\paren{{\bh x\ov\sqrt{\sigma}}-i\sqrt{\sigma}\bh p},
}
where $\lambda$ is an overall normalization, which is usually taken to be $\lambda=1/\sqrt{2}$ but we leave it as an arbitrary complex number here.
(More specifically, $\bh p$ has been defined on $L^2\fn{\mathbb R^d}$ and $\bh x$ the generator of translation on it; or equivalently, one may regard $\bh x$ to be defined on $L^2\fn{\Sigma_{(0)}}$ and $\bh p$ the generator on it.)
The dimensionality is given by $\sqbr{\bh a}=\sqbr{\lambda}$.
Note that this annihilation operator has nothing to do with the field annihilation operator in Eq.~\eqref{creation annhilation operators}.
We see that
$
\Commutator{\h a_i}{\h a_j^\dagger}
	=	2\ab{\lambda}^2\delta_{ij}\h1$ and
$\Commutator{\h a_i}{\h a_j}
	=	\Commutator{\h a_i^\dagger}{\h a_j^\dagger}=0$.
	
A coherent state is a normalizable physical state that is defined in $L^2\fn{\mathbb R^d}$ or equivalently in $L^2\fn{\Sigma_{(0)}}$ by
\al{
\bh a\Ket{\bs\alpha}
	&=	\bs\alpha\Ket{\bs\alpha},
		\label{coherent state in position momentum space}
}
where $\bs\alpha=\paren{\alpha_1,\dots,\alpha_d}$ is a $d$-vector of complex numbers of mass dimension $\sqbr{\bs\alpha}=\sqbr{\lambda}$.\footnote{
The abuse of notation should be understood that these $\alpha_i$, which are just numbers, have nothing to do with the Lorentz-friendly annihilation operator~\eqref{Lorentz-friendly annihilation operator}. Historically, the name ``coherent state'' comes from the one in field space that describes a photon coherent wave, rather than the one~\eqref{coherent state in position momentum space} in position-momentum space; see the paragraph containing Eq.~\eqref{photon coherent state}.
}
From
\al{
\lambda\paren{{{\bs x\ov\sqrt{\sigma}}+\sqrt{\sigma}\nab}}\Braket{\bs x|\bs\alpha}
	&=	\bs\alpha\Braket{\bs x|\bs\alpha},
}
we get the solution in $L^2\fn{\Sigma_{(0)}}$:
\al{
\Braket{\bs x|\bs\alpha}
	&=	{1\ov\paren{\pi\sigma}^{d\ov4}}
		e^{-{1\ov2}\paren{{\bs x\ov\sqrt{\sigma}}-\Re{\bs\alpha\ov \lambda}}^2}e^{i\Im{\bs\alpha\ov \lambda}\cdot{\bs x\ov\sqrt{\sigma}}},
			\label{coherent state wave function}
}
where we have normalized such that $\Braket{\bs\alpha|\bs\alpha}=1$ and hence $\sqbr{\Ket{\bs\alpha}}=0$.
In the momentum space $L^2\fn{\mathbb R^d}$, this becomes
\al{
\Braket{\bs p|\bs\alpha}
	&=	\paren{\pi\ov\sigma}^{d\ov4}
			e^{-i\sqrt{\sigma}\bs p\cdot\Re{\bs\alpha\ov\lambda}}e^{-{1\ov2}\paren{\sqrt{\sigma}\bs p-\Im{\bs\alpha\ov\lambda}}^2};
				\label{coherent state in momentum space}
}
see Eq.~\eqref{Gaussian basis NR}.
Physically, the real and imaginary parts of $\bs\alpha$ correspond to the center of position and momentum of the Gaussian wave packet.
Looking at Eqs.~\eqref{coherent state wave function} and \eqref{coherent state in momentum space}, it is rather mysterious why the wave functions take such particular forms as functions of complex numbers $\bs\alpha$. We will shed some light on this point in Sec.~\ref{Lorentz-invariant wave packet section}.

Comparing Eq.~\eqref{coherent state wave function} with Eq.~\eqref{Gaussian wave function}, we see that the Gaussian wave-packet state is indeed a coherent state in $L^2\fn{\mathbb R^d}$ or equivalently in $L^2\fn{\Sigma_{(0)}}$:
\al{
\Ket{\bs X,\bs P}
	&=	\Ket{\lambda\,\Bigl({\bs X\ov\sqrt{\sigma}}+i\sqrt{\sigma}\bs P\Bigr)}.
}
By taking $\lambda=\sqrt{\sigma}$ and $-i/\sqrt{\sigma}$, we may write
\al{
\Ket{\bs X,\bs P}
	&=	\Ket{\bs X+i\sigma\bs P}_{\lambda=\sqrt{\sigma}}
	=	\Ket{\bs P-i{\bs X\ov\sigma}}_{\lambda=-{i\ov\sqrt{\sigma}}}.
	\label{Gaussian by coherent}
}
Now we see that the completeness relation~\eqref{completeness of Gaussian basis} is equivalent to the completeness of the coherent states in $L^2\fn{\mathbb R^d}$ or equivalently in $L^2\fn{\Sigma_{(0)}}$:
\al{
{1\ov\paren{2\pi\ab{\lambda}^2}^d}\int\df^{2d}\bs\alpha\,\Ket{\bs\alpha}\Bra{\bs\alpha}
	&=	\h1,
}
where
\al{
\df^{2d}\bs\alpha
	&=	\prod_{i=1}^d\df\Re\alpha_i\,\df\Im\alpha_i
	=	\bigwedge_{j=1}^d\paren{{i\ov2}\df\alpha_j\wedge\df\alpha_j^*}.
}
We list some more usable facts in Appendix~\ref{coherent appendix}.

We comment that the coherent state in the position-momentum space~\eqref{coherent state wave function} or \eqref{coherent state in momentum space} should not be confused with the coherent state in the (photon) \emph{field space}, used in quantum optics~\cite{Sudarshan1963,Glauber1963}, for a \emph{fixed} wavenumber vector $\bs k$ (and hence with the fixed wavelength $2\pi/\ab{\bs k}$):
\al{
\Ket{ z}_{\bs k}
	&=	e^{-{\ab{ z}^2\ov2}}e^{ z\wh a_{\bs k}^\dagger}\Ket{0},
		\label{photon coherent state}
}
where $\wh a_{\bs k}^\dagger$ is the creation operator in the sense of Eq.~\eqref{creation annhilation operators} (but with a box normalization $\Commutator{\wh a_{\bs k}}{\wh a_{\bs k'}^\dagger}=\delta_{\bs k,\bs k'}$) and we have taken $\lambda=1/\sqrt{2}$.

\section{Lorentz-invariant wave packet}\label{Lorentz-invariant wave packet section}
From the form of the Gaussian wave-packet state~\eqref{Gaussian basis}, it is tempting to generalize it into a Lorentz invariant form:
\al{
\Braket{\bs p|\bs X,\bs P}
	&\propto
		e^{-i\bs p\cdot\bs X-{\sigma\ov2}\paren{\bs p-\bs P}^2}
	\to	e^{-ip\cdot X-{\sigma\ov2}\paren{p-P}^2}
	=		e^{\sigma m^2}e^{-ip\cdot\paren{X+i\sigma P}},
	\label{a possible generalization}
}
where we have used the on-shell condition $p^2=P^2=-m^2$.
As we have seen in Eq.~\eqref{Gaussian by coherent}, the Gaussian wave-packet state $\Ket{\bs X,\bs P}$ is nothing but a position-momentum coherent state. For the coherent state, it is rather mysterious why the real and imaginary parts of the complex numbers $\bs\alpha$ appear in the forms~\eqref{coherent state wave function} and \eqref{coherent state in momentum space}. It is remarkable that the Lorentz invariant generalization~\eqref{a possible generalization} has the seemingly holomorphic dependence on the $D$ complex variables $X+i\sigma P$ if one generalizes $P$ to be off-shell.\footnote{
As we will discuss below, the generalization of $P$ to an off-shell momentum is straightforward so far as $P$ is timelike and future-oriented. We leave further generalization for future study.
}

Motivated by this fact, we define the following Lorentz-invariant wave-packet state in $L^2\fn{\mathbb R^d}$:\footnote{
To be precise, the state $\dK{\sigma;X,\bs P}$ is Lorentz covariant and the wave function~\eqref{Lorentzian packet with p} is Lorentz invariant.
}
\al{
\dBraket{\Braket{\bs p|\sigma;X,\bs P}}
	&:=	N_\sigma e^{-ip\cdot\paren{X+i\sigma P}}\nn
	&=	N_\sigma e^{iE_{\bs p}X^0-i\bs p\cdot\bs X}e^{-\sigma\paren{E_{\bs p}E_{\bs P}-\bs p\cdot\bs P}},
		\label{Lorentzian packet with p}
}
where $N_\sigma$ is a normalization constant.
Given the reference frame $x$, this wave packet is centered near $\bs x=\bs X$ and $\bs p=\bs P$ at time $x^0=X^0$.
As said above, one might want to regard the state $\dK{\sigma;X,\bs P}$ as a basis of $\Sigma_{(X^0)}$ in the interaction picture but we will see that, in terms of the wave function~\eqref{Lorentz-invariant wave function} in $\Sigma_{(x^0)}$, $X^0$ is mere a parameter that specifies the wave packet $\dK{\sigma;X,\bs P}$, while the time is specified by $x^0$.
Also, we will continue to abbreviate $\sigma$ to write $\dKet{\Ket{X,\bs P}}$ unless otherwise stated.

As an illustration of more general computation spelled out in Appendix~\ref{basic integral},
we will show in Sec.~\ref{normalization section} that the normalization
\al{
\Abb{\dKet{\Ket{X,\bs P}}}^2
	&=	1,
	\label{normalization of LIP}
}
in $L^2\fn{\mathbb R^d}$, is realized by
\al{
N_\sigma
	&=	\paren{\sigma\ov\pi}^{d-1\ov4}\paren{K_{d-1\ov2}\fn{2\sigma m^2}}^{-{1\ov2}},
	\label{normalization constant}
}
where $K$ is the modified Bessel function of the second kind.
With this normalization, mass dimensions are
$
\Sqbr{\dKet{\Ket{X,\bs P}}}
	=	0$ and
$
\sqbr{N_\sigma}
	=	-{d-1\ov2}$.

We comment on possible generalizations of $P$ to be off-shell.
If we make $P$ off-shell in the first line in Eq.~\eqref{Lorentzian packet with p}, it becomes divergent for $\ab{\bs p}\to\infty$ when $p\cdot P>0$, namely, when
\al{
P^0	&<	\bs v\cdot\bs P,
	\label{divergent condition}
}
where $\bs v:=\bs p/E_{\bs p}$ with $\ab{\bs v}<1$.
Therefore, the generalization of $P$ to off-shell would be safe so long as $P$ is timelike and future-oriented, in which case the condition~\eqref{divergent condition} is never met.
(This is the case too if we let $P$ be off-shell in $e^{-ip\cdot X-{\sigma\ov2}\paren{p-P}^2}=e^{-ip\cdot X}
		e^{{\sigma\ov2}\paren{m^2-P^2}+\sigma p\cdot P}$, though the limit of super-heavy ``off-shell mass'' $-P^2\to\infty$ diverges.)

\subsection{Normalization}\label{normalization section}
We compute the norm on $L^2\fn{\mathbb R^d}$:
\al{
\Abb{\dKet{\Ket{X,\bs P}}}^2
	&=	\int{\df^d\bs p\ov2E_{\bs p}}\dBraket{\Braket{X,\bs P|\bs p}}\dBraket{\Braket{\bs p|X,\bs P}}
	=	\ab{N_\sigma}^2\int{\df^d\bs p\ov2E_{\bs p}}e^{2\sigma p\cdot P}\nn
	&=	\ab{N_\sigma}^2\int\df^Dp\,\theta\fn{p^0}\delta\fn{p^2+m^2}e^{2\sigma p\cdot P},
}
where we let $p$ be off-shell in the last line to make the Lorentz invariance manifest.
As $P$ is on-shell, we may always find a Lorentz transformation $\Lambda\fn{P}$ to its rest frame $\wt P:=\paren{m,\bs 0}$ such that $\Lambda P=\wt P$.
Then we change the integration variable to $\wt p:=\Lambda p$.
Using the Lorentz invariance of the integration measure etc.\ as well as $\paren{\Lambda^{-1}\wt p}\cdot P=\wt p\cdot\Lambda P=\wt p\cdot\wt P=-\wt p^0m$, we get
\al{
\Abb{\dKet{\Ket{X,\bs P}}}^2
	&=	\ab{N_\sigma}^2\int\df^D\wt p\,\theta\fn{\wt p^0}\delta\fn{\wt p^2+m^2}
e^{-2\sigma \wt p^0m}
	=	\ab{N_\sigma}^2\int{\df^d\bs p\ov2E_{\bs p}}
		e^{-2\sigma m\sqrt{m^2+\bs p^2}}\nn
	&=	\ab{N_\sigma}^2\Omega_{d-1}\int_m^\infty{\paren{E^2-m^2}^{d-2\ov2}E\,\df E\ov2E}
		e^{-2\sigma mE}
	=	\ab{N_\sigma}^2\paren{\pi\ov\sigma}^{d-1\ov2}K_{d-1\ov2}\fn{2\sigma m^2},
}
where $\Omega_{d-1}=2\pi^{d\ov2}/\Gamma\fn{d\ov2}$ is the area of a unit $\paren{d-1}$-sphere.
We see that the normalization $\abb{\Ket{X,\bs P}}^2=1$ is realized by Eq.~\eqref{normalization constant}.
One can also check that this result is consistent with the master formula~\eqref{master formula} with $\Xi\to2\sigma P$ and hence $\abb\Xi=\sqrt{-\Xi^2}\to2\sigma m$.

In the following, we list several limits for the reader's ease.
First,
\al{
K_n\fn{z}
	&=	\begin{cases}
		{2^{n-1}\paren{n-1}!\ov z^{n}}+\Or{1\ov z^{n-1}}
			&	(z\to0,\,n>0),\\
		e^{-z}\sqbr{\sqrt{\pi\ov2z}+\Or{1\ov z^{3/2}}}
			&	(z\to\infty),
		\end{cases}
}
and in the limits $\sigma m^2\to0$ and $\infty$, we get, respectively,
\al{
N_\sigma
	&\to	\begin{cases}
			\sqrt{2\ov\paren{d-3\ov2}!}
			\paren{\sigma m\ov\sqrt{\pi}}^{d-1\ov2}
				&	(\sigma m^2\to0),\\
			\sqrt{2m}\paren{\sigma\ov\pi}^{d\ov4}e^{\sigma m^2}
				&	(\sigma m^2\to\infty).
			\end{cases}
		\label{limit of C}
}
Here one might find it curious that a plane-wave limit $\sigma\to\infty$ is equivalent to a non-relativistic  limit $m\to\infty$, and a particle limit $\sigma\to0$ to an ultra-relativistic limit $m\to0$.

The non-relativistic limit $m\to\infty$ of the Lorentz-invariant wave packet~\eqref{Lorentzian packet with p} comes back to the Gaussian form~\eqref{Gaussian basis} up to the factor $\sqrt{2m}$,
\al{
\dBraket{\Braket{\bs p|X,\bs P}}
	&\to	\sqrt{2m}\paren{\sigma\ov\pi}^{d\ov4}e^{iE_{\bs p}X^0-i\bs p\cdot\bs X}
			e^{-{\sigma\ov2}\paren{\bs p-\bs P}^2},
				\label{non relativistic limit}
}
where $E_{\bs p}=m+{\bs p^2\ov2m}+\cdots$, we have used the limit~\eqref{limit of C}, and have neglected $\Or{m^{-2}}$ terms in the last exponent in Eq.~\eqref{non relativistic limit}.
In the ultra-relativistic limit $m\to0$, we get
\al{
\dBraket{\Braket{\bs p|X,\bs P}}
	&\to	\sqrt{2\ov\paren{d-3\ov2}!}
			\paren{\sigma m\ov\sqrt{\pi}}^{d-1\ov2}
			e^{-\ab{\bs p}\br{
				\sigma\ab{\bs P}\paren{
					1-\cos\theta_{\bs P}}
				-i\paren{X^0-\ab{\bs X}\cos\theta_{\bs X}}
				}},
}
where $\cos\theta_{\bs X}:={\bs p\cdot\bs X}/\ab{\bs p}\ab{\bs X}$ and $\cos\theta_{\bs P}:={\bs p\cdot\bs P}/\ab{\bs p}\ab{\bs P}$.
In this limit, the original Gaussian suppression is made weaker. Especially along the direction of $\bs P$, $\cos\theta_{\bs P}=1$, there is no suppression for a large momentum $\ab{\bs p}\to\infty$.
This is the main obstacle of having a Lorentz-invariant wave packet for a massless particle.

\subsection{Inner product}
Let us compute the inner product of two Lorentz invariant wave packets: $\dBK{\sigma;X,\bs P|\sigma';X',\bs P'}$ on $L^2\fn{\mathbb R^d}$.
Here and hereafter, a prime symbol $'$ never denotes a derivative.

Motivated by the coherent states in the position-momentum space~\eqref{Gaussian by coherent}, we define the following complex variables
\al{
Z^\mu\fn{\sigma}
	&:=	X^\mu+i\sigma P^\mu,\\
\Pi^\mu\fn{\sigma}
	&:=	P^\mu-i{X^\mu\ov\sigma},
}
which are related by $Z^\mu=i\sigma\Pi^\mu$.
We define the Lorentz invariant analog of the coherent state~\eqref{Gaussian by coherent}:\footnote{
Abuse of notation is understood: Always the first and second definitions in Eq.~\eqref{Loreintz-invariant coherent states} are used for the arguments $Z$ and $\Pi$, respectively.
}
\al{
\dKet{\Ket{Z\fn{\sigma}}}
	&:=	\dKet{\Ket{\sigma;X,\bs P}},&
\dKet{\Ket{\Pi\fn{\sigma}}}
	&:=	\dKet{\Ket{\sigma;X,\bs P}}.
	\label{Loreintz-invariant coherent states}
}
We see that
\al{
\dBraket{\Braket{\bs p|Z\fn{\sigma}}}
	&=	N_\sigma e^{-ip\cdot Z\paren{\sigma}},&
\dBraket{\Braket{\bs p|\Pi\fn{\sigma}}}
	&=	N_\sigma e^{\sigma p\cdot\Pi\paren{\sigma}}.
}
For later use, we define a ``norm'' of an arbitrary complex $D$-vector $\Xi$:
\al{
\abb{\Xi}
	&:=	\sqrt{-\Xi^2}
	=	\sqrt{\pn{\Xi^0}^2-\bs\Xi^2},
}
which is not necessarily positive nor even a real number.\footnote{
The abuse of notation should be understood: This has nothing to do with the norm of a state vector in the Hilbert space such as in Eq.~\eqref{normalization of LIP}.
}
Now we may write
\al{
\dBraket{\Braket{\sigma;X,\bs P|\sigma';X',\bs P'}}
	=	\dBraket{\Braket{\Pi\fn{\sigma}|\Pi'\fn{\sigma'}}}
	=	N_\sigma^* N_{\sigma'}
		\int{\df^d\bs p\ov2E_{\bs p}}e^{p\cdot\paren{\sigma\Pi^*\paren{\sigma}+\sigma'\Pi'\paren{\sigma'}}}.
			\label{inner product}
}

To compute this, it is convenient to define the master integral:
\al{
\mc I\fn{\Xi}
	&:=	\int{\df^d\bs u\ov2u^0}e^{u\cdot\Xi}
	=	\int\df^Du\,\delta\fn{u^2+1}\theta\fn{u^0}e^{u\cdot\Xi},
		\label{master integral}
}
where $\Xi$ is a dimensionless complex $D$-vector and the $D$-vector $u$ ($=p/m$) is on-shell and off-shell for the first and second integrals, respectively.
In Appendix~\ref{basic integral}, we present a detailed evaluation of the integral, and the result is
\al{
\mc I\fn{\Xi}
	&=	\pn{2\pi}^{d-1\ov2}
		{K_{d-1\ov2}\Fn{\abb{\Xi}}\ov
		\abb{\Xi}^{{d-1\ov2}}},
		\label{master formula}
}
which is valid when $\Re\Xi$ is timelike $\pn{\Re\Xi}^2<0$ and future-oriented $\Re\Xi^0>0$.
This also implies that
\al{
\mc I\fn{2\sigma mP}
	=	{1\ov m^{d-1}N_\sigma^2},
}
with $N_\sigma$ is given in Eq.~\eqref{normalization constant}.

The integral in the inner product~\eqref{inner product} corresponds to
\al{
\Xi
	&=	m\pn{\sigma\Pi^*\fn{\sigma}+\sigma'\Pi'\fn{\sigma'}}
	=	im\paren{Z^*\fn{\sigma}-Z'\fn{\sigma'}},
}
that is,
\al{
\Re\Xi
	&=	m\pn{\sigma P+\sigma'P'},&
\Im\Xi
	&=	m\pn{X-X'}.
}
From
\al{
\paren{\sigma P+\sigma'P'}^2
	&=	-\paren{\sigma^2+\sigma^{\pr2}}m^2+2\sigma\sigma'P\cdot P'\nn
	&\leq
		-\paren{\sigma^2+\sigma^{\pr2}}m^2
		+2\sigma\sigma'\paren{-\sqrt{\ab{\bs P}^2+m^2}\sqrt{\ab{\bs P'}^2+m^2}+\ab{\bs P}\ab{\bs P'}}
	<	0,
}
we see that $\Re\Xi$ is always timelike. Therefore we may use the result~\eqref{master formula}:
\al{
\dBK{\sigma;X,\bs P|\sigma';X',\bs P'}
	&=	N_\sigma^*N_{\sigma'}
		\paren{2\pi}^{d-1\ov2}m^{d-1}
		{K_{d-1\ov2}\fn{\abb\Xi}\ov
		\abb\Xi^{{d-1\ov2}}}\nn
	&=	{\paren{2\sqrt{\sigma\sigma'}m^2}^{d-1\ov2}\ov
			\paren{K_{d-1\ov2}\fn{2\sigma m^2}K_{d-1\ov2}\fn{2\sigma' m^2}}^{1/2}
			}
		{K_{d-1\ov2}\fn{\abb\Xi}\ov
		\abb\Xi^{{d-1\ov2}}},
}
where in various notations,
\al{
\abb\Xi^2
	=	-\Xi^2
	&=	-m^2\paren{\sigma\Pi^*+\sigma'\Pi'}^2&
	&=	m^2\paren{Z^*-Z'}^2\nn
	&=	-m^2\Paren{\sigma P+\sigma'P'+i\paren{X-X'}}^2&
	&=	m^2\Paren{\paren{X-X'}-i\paren{\sigma P+\sigma'P'}}^2.
}
Especially when $\sigma=\sigma'$, we have
\al{
\dBraket{\Braket{X,\bs P|X',\bs P'}}
	&=	{\paren{2\sigma m^2}^{d-1\ov2}\ov
			K_{d-1\ov2}\fn{2\sigma m^2}
			}
		{K_{d-1\ov2}\fn{\abb\Xi}\ov
		\abb\Xi^{{d-1\ov2}}},
}
where
\al{
\abb\Xi^2
	&=	-m^2\sigma^2\paren{\Pi^*+\Pi'}^2&
	&=	m^2\paren{Z^*-Z'}^2\nn
	&=	-m^2\Paren{\sigma\paren{P+P'}+i\paren{X-X'}}^2&
	&=	m^2\Paren{\paren{X-X'}-i\sigma\paren{P+P'}}^2.
}

\subsection{Wave function}\label{wave function section}
Let us compute the wave function for the Lorentz-invariant wave-packet state on $L^2\fn{\Sigma_{(x^0)}}$:
\al{
\dBraket{\Braket{x|X,\bs P}}
	&=	\int{\df^d\bs p\ov2E_{\bs p}}\dBraket{\Braket{x|\bs p}}\dBraket{\Braket{\bs p|X,\bs P}}
	=	{N_\sigma\ov\paren{2\pi}^{d\ov2}}\int{\df^d\bs p\ov2E_{\bs p}}e^{p\cdot\paren{\sigma P+i\paren{x-X}}}.
}
Comparing with Eq.~\eqref{master integral}, we see the following correspondence:
\al{
\Xi
	=	m\Pn{\sigma P+i\paren{x-X}}.
		\label{Xi for wave function}
}
Obviously $\pn{\Re\Xi}^2=-\sigma^2m^4<0$, and we may use Eq.~\eqref{master formula}:
\al{
\dBraket{\Braket{x|X,\bs P}}
	&=	
		{N_\sigma m^{d-1}\ov\sqrt{2\pi}}
		{K_{d-1\ov2}\fn{\abb\Xi}\ov
		\abb\Xi^{d-1\ov2}}
	=	
		{1\ov\paren{4\pi\sigma}^{d-1\ov4}}
		{\paren{2\sigma m^2}^{d-1\ov2}
			\ov\sqrt{2\pi K_{d-1\ov2}\fn{2\sigma m^2}}}
		{K_{d-1\ov2}\fn{\abb\Xi}\ov
		\abb\Xi^{{d-1\ov2}}},
		\label{Lorentz-invariant wave function}
}
where
\al{
\abb\Xi
	:=	\sqrt{-\Xi^2}
	=	m\sqrt{\sigma^2m^2+\paren{x-X}^2-2i\sigma P\cdot\paren{x-X}}.
	\label{argument Xi}
}
The explicit form of the wave function~\eqref{Lorentz-invariant wave function} is one of our main results.
This reduces to the earlier one in Refs.~\cite{Naumov:2010um,Naumov:2020yyv} when $d=3$ in the $X\to0$ limit and may be interpreted as its spacetime translation by $X$.
We note that there is no branch-cut ambiguity for the argument~\eqref{argument Xi} as long as $m>0$; see the last paragraph in Appendix~\ref{Master integral section}.


Hereafter, we examine various characteristics of the above wave function.
Firstly, along the line $x=X+Ps$ corresponding to the particle trajectory, with $s$ being a real parameter, we get $\Xi=mP\pn{\sigma+is}$ and hence
$\abb\Xi
	=	m^2\pn{\sigma+is}$.
For a point sufficiently apart from $X$ along this trajectory, namely for $s\to\pm\infty$, we get
\al{
\dBraket{\Braket{x|X,\bs P}}
	&\to
		{\paren{\sigma\ov\pi}^{d-1\ov4}
			\ov2m\sqrt{K_{d-1\ov2}\fn{2\sigma m^2}}}
		{e^{-m^2\pn{\sigma+is}}\ov\pn{\sigma+is}^{d\ov2}}.
}
We see that the wave function is not suppressed along the direction of $P$: There is no exponential suppression for $\ab s\to\infty$, while the apparent power suppression $\propto\ab s^{-d}$ for $\ab{\dBraket{\Braket{x|X,\bs P}}}^2$ is merely due to the broadening of the width of Gaussian wave packet in $d$-spatial dimensions (for a normalized wave packet, the height of center becomes lower and lower when the width is more and more broadened), as we will soon see below.

Secondly, let us further consider a point slightly away from this trajectory, $x=X+Ps+\epsilon$, where $\epsilon$ is a small spacelike $D$-vector: $\epsilon^2>0$.
(Here, for each $s$, a point on the trajectory $X+Ps$ is specified, and we parametrize the spacelike hyperplane containing that point by $\epsilon$.)
Then in the limit $\ab s\to\infty$,
\al{
\dBraket{\Braket{x|X,\bs P}}
	&\to	
			{\paren{\sigma\ov\pi}^{d-1\ov4}
			\ov2m\sqrt{K_{d-1\ov2}\fn{2\sigma m^2}}}
			{e^{-\sigma m^2-im^2s}\ov \pn{\sigma+is}^{d\ov2}}
			\exp\fn{-{\sigma\pn{\epsilon^2+\pn{{P\ov m}\cdot\epsilon}^2}\ov
				2\pn{s^2+\sigma^2}}
			+iP\cdot\epsilon},
			\label{limit of wave function}
}
where we have discarded $\Or{\epsilon^2}$ and $\Or{\epsilon^3}$ terms in the imaginary and real parts of the exponent, respectively, and $\Or{\epsilon}$ terms in other places.
We observe the plane-wave behavior, $e^{iP\cdot\epsilon}$, and we obtain the Gaussian suppression factor: $\exp\Sqbr{-{\sigma\ov
				2\pn{s^2+\sigma^2}}
		\Pn{\epsilon^2+\pn{{P\ov m}\cdot\epsilon}^2}
		}$.
It is noteworthy that the more we go along the trajectory $x=X+sP$ (namely the larger the $\ab s$ is), the larger the spatial width-squared $\sim\pn{s^2+\sigma^2}/\sigma$ of this Gaussian factor becomes.

For a wave-packet scattering, we may parametrize each of the incoming waves, $a$, and of the outgoing ones, $b$, such that the scattering occurs (i.e.\ the wave packets overlap) around finite region $\ab{s_a}\sim\ab{s_b}<\infty$.
If the scattering occurs within a large time interval, the in and out asymptotic states are given by $s_a\to-\infty$ and $s_b\to\infty$, respectively.
In such a case, we may approximate an in-coming/out-going wave packet by the near plane wave~\eqref{limit of wave function} better and better, whereas they still interact as wave packets rather than plane waves.

Thirdly, for the plane-wave expansion with large $\sigma$, the argument becomes\footnote{
On the other hand, when we take the non-relativistic limit $m\to\infty$ first, we get
$
\abb\Xi
	=	\sigma m^2
		+im\pn{x^0-X^0}
			-i\bs P\cdot\pn{\bs x-\bs X}
			+{\pn{\bs x-\bs X}^2\ov2\sigma}
			+\Or{1\ov m}
$.
}
\al{
\abb\Xi
	&=	\sigma m^2
		-iP\cdot\paren{x-X}
			+{\paren{x-X}^2\ov2\sigma}
			+{\pn{P\cdot\pn{x-X}}^2\ov2\sigma m^2}
			+\Or{1\ov\sigma^2},
			\label{expansion of argument}
}
where we have taken up to the order of leading non-trivial real part, and the wave function becomes
\al{
\dBraket{\Braket{x|X,\bs P}}
	&\simeq
		{\paren{\sigma\ov\pi}^{d\ov4}
			\ov\sqrt2\paren{\sigma m}^{d-{1\ov2}}}
		e^{
			iP\cdot\paren{x-X}
			-{1\ov2\sigma}\paren{x-X}^2
			-{\pn{P\cdot\pn{x-X}}^2\ov2\sigma m^2}
			}
		.\label{plane-wave limit of Lorentz invariant wave packet}
}
The corresponding probability density is
\al{
\ab{\dB{x}\sqrt{2E_{\bh p}}\dK{X,\bs P}}^2
	&\simeq
		{E_{\bs P}\paren{\sigma\ov\pi}^{d\ov2}
			\ov\paren{\sigma m}^{{d\ov2}-1}}
		e^{
			-{1\ov\sigma}\paren{x-X}^2
			-{\pn{P\cdot\pn{x-X}}^2\ov\sigma m^2}
			},
}
where we have used the completeness~\eqref{completeness in Lorentz-friendly x}.

In particular on the line $x=X+Ps$, with $s$ being a real parameter, the quadratic terms of $s$ cancel out in the exponent in Eq.~\eqref{plane-wave limit of Lorentz invariant wave packet}:
\al{
\dBraket{\Braket{x|X,\bs P}}
	&\propto	{\paren{\sigma\ov\pi}^{d\ov4}
			\ov\sqrt2\paren{\sigma m}^{d-{1\ov2}}}e^{-im^2s}.
}
As promised, we have confirmed that the wave function does not receive the Gaussian suppression along this particle trajectory. We stress that in this sense, the Lorentz-invariant wave packet is not localized in time, just as the Gaussian wave packet reviewed in Sec.~\ref{Gaussian basis section}.

If we further take the non-relativistic limit in the exponent of the large-$\sigma$ expansion~\eqref{plane-wave limit of Lorentz invariant wave packet}, it becomes
\al{
iP\cdot\pn{x-X}
-{\pn{\bs x-\bs X-\bs V\fn{x^0-X^0}}^2\ov2\sigma}
-{\pn{\bs V\cdot\pn{\bs x-\bs X}}^2\ov2\sigma}
+\Or{\ab{\bs V}^3},
	\label{NR limit of exponent}
}
where $P^0=m+{m\ov2}\bs V^2+\cdots$. Comparing with the Gaussian wave packet~\eqref{closed form of Gaussian wave function}, we see that the extra suppression factor
\al{
\exp\fn{-{\pn{\bs V\cdot\pn{\bs x-\bs X}}^2\ov2\sigma}}
}
appears from the Lorentz-invariant wave packet, and the center of the Lorentz-invariant wave packet departs from the particle trajectory $\bs X-\bs V\pn{x^0-X^0}$ of the Gaussian wave packet~\eqref{closed form of Gaussian wave function}.

Finally, in the particle/ultra-relativistic limit $\sigma m^2\to0$, we get 
\al{
\dBK{x|X,\bs P}
	&\to	
		{\pn{\sigma m^3}^{d-1\ov2}\ov\sqrt{\pn{d-3\ov2}!}}{K_{d-1\ov2}\fn{\abb\Xi}\ov\abb\Xi^{d-1\ov2}},
}
where the argument goes to
$
\abb\Xi
	\to	m\sqrt{\paren{x-X}^2
			-2i\sigma P\cdot\paren{x-X}}$.
If we further take the relativistic limit $m\to0$, we get\footnote{
On the other hand, first taking the particle limit $\sigma\to0$ is tricky due to the branch cut:
When $x$ is located at a spacelike distance from $X$, namely $\paren{x-X}^2>0$,
\als{
\abb\Xi
	&\to	m\sqrt{\paren{x-X}^2}\paren{
				1-i{\sigma P\cdot\paren{x-X}\ov\paren{x-X}^2}+\cdots
				};
}
when timelike, $\paren{x-X}^2<0$,
\als{
\abb\Xi
	&\to	-im\sgn\Fn{P\cdot\paren{x-X}}\sqrt{-\paren{x-X}^2}\paren{
			1
			+i{\sigma P\cdot\paren{x-X}\ov-\paren{x-X}^2}+\cdots
			};
}
and when lightlike $\paren{x-X}^2=0$,
\als{
\abb\Xi
	&\to	m\sqrt{-2i\sigma P\cdot\paren{x-X}}
			\paren{1+i{\sigma m^2\ov4P\cdot\paren{x-X}}+\cdots}.
}
}
\al{
\dBK{x|X,\bs P}
	&\to
		{\sqrt{\Gamma\fn{d-1\ov2}}\ov2}
		\pn{2\sigma m\ov\paren{x-X}^2
			-2i\sigma P\cdot\paren{x-X}}^{d-1\ov2}.
}


\section{Uncertainty relations}\label{uncertainty section}
We show how the uncertainty relation changes for the Lorentz-invariant wave packet.
We study the momentum and position uncertainties in the first two subsections and then discuss the uncertainty relation in the next. Lastly, we comment on the time-energy uncertainty.

\subsection{Momentum (co)variance}
We want to compute the momentum expectation value $\Braket{\h p^\mu}$ and its (co)variance (recall that we have been taking all the momenta on-shell, and hence $\h p^0=E_{\bh p}=\sqrt{m^2+\bh p^2}$):
\al{
\Braket{
	\Pn{\h p^\mu-\Braket{\h p^\mu}}
	\Pn{\h p^\nu-\Braket{\h p^\nu}}
	}
	&=	\Braket{\h p^\mu\h p^\nu}-\Braket{\h p^\mu}\Braket{\h p^\nu},
}
where for any operator $\h{\mc O}$, we write the expectation value with respect to $\dK{X,\bs P}$ as
\al{
\braket{\h{\mc O}}
	&:=	\dB{X,\bs P}\h{\mc O}\dK{X,\bs P}.
	\label{expectation value}
}

Since we identify the Schr\"odinger, Heisenberg, and interaction pictures at $x^0=0$, the expectation value~\eqref{expectation value} corresponds to a measurement on the spacelike hyperplane $\Sigma_{(0)}$.
A measurement on a different time slice $\Sigma_{(x^0)}$ is given by
\al{
\braket{\h{\mc O}_\tx{H}\fn{x^0}}
	&:=	\dB{X,\bs P}e^{i\h Hx^0}\h{\mc O}e^{-i\h Hx^0}\dK{X,\bs P}.
}
As we are only considering free propagation of the waves, $\h H=\h H_\tx{free}$, this is the same as
\al{
\braket{\h{\mc O}_\tx{I}\fn{x^0}}
	:=	\dB{X,\bs P}e^{i\h H_\tx{free}x^0}\h{\mc O}e^{-i\h H_\tx{free}x^0}\dK{X,\bs P}
}
for our application. In particular when $\h O$ only contains momentum operators such that $\commutator{\h O}{\h H_\tx{free}}=0$, the expectation value becomes time independent $\Braket{\h O}=\Braket{\h O_\tx{I}\fn{x^0}}$.

First, we write
\al{
\Braket{\h p^{\mu_1}\cdots\h p^{\mu_n}}
	&=	\int{\df^d\bs p\ov2E_{\bs p}}\Ab{\dBK{\bs p|X,\bs P}}^2p^{\mu_1}\cdots p^{\mu_n}
	=	N_\sigma^2m^{d-1+n}\int{\df^d\bs u\ov2u^0}e^{2\sigma mu\cdot P}u^{\mu_1}\cdots u^{\mu_n},
}
where $u=p/m$ is the $D$-velocity with $u^0=\sqrt{1+\bs u^2}$.
To compute the above, we take derivatives of the master integral~\eqref{master formula}:
\al{
\mc I^{\mu_1\dots\mu_n}\fn{\Xi}
	&:=	{\p^n\ov\p\Xi_{\mu_1}\cdots\p\Xi_{\mu_n}}\mc I\fn{\Xi}
	=	\int{\df^d\bs u\ov2u^0}e^{u\cdot\Xi}u^{\mu_1}\cdots u^{\mu_n},
}
where $\Xi$ is off-shell. Once this is obtained, we may substitute $\Xi=2\sigma mP$, which is ``on-shell'', $\abb{\Xi}=\sqrt{-\Xi^2}=2\sigma m^2$.

From Eqs.~\eqref{I mu} and \eqref{I mu nu} in Appendix~\ref{basic integral}, we can readout
\al{
\Braket{\h p^\mu}
	&=	m{\mc I^\mu\fn{2\sigma mP}\ov\mc I\fn{2\sigma mP}}
	=	
		{K_{d+1\ov2}\fn{\abb{\Xi}}\ov K_{d-1\ov2}\fn{\abb{\Xi}}}
		P^\mu,
			\label{p mu}\\
\Braket{\h p^\mu\h p^\nu}
	&=	m^2{\mc I^{\mu\nu}\fn{2\sigma mP}\ov\mc I\fn{2\sigma mP}}
	=	\paren{
			{\eta^{\mu\nu}\ov2\sigma}{K_{d+1\ov2}\fn{\abb{\Xi}}\ov K_{d-1\ov2}\fn{\abb{\Xi}}}
			+P^\mu P^\nu
				{K_{d+3\ov2}\fn{\abb{\Xi}}\ov K_{d-1\ov2}\fn{\abb{\Xi}}}
			},
		\label{p mu nu}
}
where $\abb{\Xi}=2\sigma m^2$, and hence
\al{
\Braket{\h p^\mu\h p^\nu}-\Braket{\h p^\mu}\Braket{\h p^\nu}
	&=	{\eta^{\mu\nu}\ov2\sigma}{K_{d+1\ov2}\fn{\abb\Xi}\ov K_{d-1\ov2}\fn{\abb\Xi}}
		+P^\mu P^\nu
				\pn{
					{K_{d+3\ov2}\fn{\abb\Xi}\ov K_{d-1\ov2}\fn{\abb\Xi}}
					-\paren{K_{d+1\ov2}\fn{\abb\Xi}\ov K_{d-1\ov2}\fn{\abb\Xi}}^2
					}.
					\label{relativistic covariance}
}
Note that contraction of Eq.~\eqref{p mu nu} with the flat metric $\eta_{\mu\nu}$ gives $\eta_{\mu\nu}\Braket{\h p^\mu\h p^\nu}=-m^2$ as it should, due to the Bessel identity~\eqref{Bessel identity}. We see that for a fixed $\sigma$ and $m$, the (co)variance~\eqref{relativistic covariance} becomes larger and larger for $\ab{\bs P}\to\infty$ due to the second term.
Furthermore, even the off-diagonal covariance for $\mu\neq\nu$ is non-zero. This is due to the fact that, with $\bs P\neq0$, the Lorentz-invariant wave packet is boosted and is not spherically symmetric in the momentum space, unlike the Gaussian wave packet~\eqref{Gaussian basis}.
The above results agree with Eqs.~(4.4) and (4.5) in Ref.~\cite{Kaiser:1977ys}.

From Eq.~\eqref{p mu}, we obtain
\al{
-\Braket{\h p}^2
	&=	m^2\pn{K_{d+1\ov2}\fn{\abb\Xi}\ov K_{d-1\ov2}\fn{\abb\Xi}}^2
	\geq m^2,
}
where the equality holds in the plane-wave/non-relativistic limit $\abb\Xi=2\sigma m^2\to\infty$.
It is curious that the mass constructed from the expectation value of $D$-momentum $\Braket{p^\mu}$ becomes larger than the ``intrinsic'' mass $m$, no matter whether the particle is at rest $\bs P=0$ or not. This fact has been pointed out in Ref.~\cite{Naumov:2020yyv}.

In the plane-wave/non-relativistic expansion for large $\abb\Xi=2\sigma m^2$, we get
\al{
\Braket{\h p^\mu}
	&=	\paren{1+{d\ov2\abb\Xi}+{d\pn{d-2}\ov8\abb\Xi^2}
			+\cdots}P^\mu,\\
\Braket{\h p^\mu\h p^\nu}
	&=	{\eta^{\mu\nu}\ov2\sigma}\paren{1+{d\ov2\abb\Xi}+{d\pn{d-2}\ov8\abb\Xi^2}+\cdots}
			+P^\mu P^\nu\paren{1+{d+1\ov\abb\Xi}+{d\pn{d+1}\ov2\abb\Xi^2}+\cdots},
		\label{pp expectation value}
}
and hence,
\al{
\Braket{\h p^\mu\h p^\nu}-\Braket{\h p^\mu}\Braket{\h p^\nu}
	&=	{\eta^{\mu\nu}\ov2\sigma}
				\paren{1+{d\ov2\abb\Xi}+{d\pn{d-2}\ov8\abb\Xi^2}+\cdots}
			+P^\mu P^\nu\paren{{1\ov\abb\Xi}+{d\ov\abb\Xi^2}+\cdots},
				\label{covariance pp}
}
where the dots denote terms of order $\abb\Xi^{-3}$.
As a cross-check, we can derive from Eq.~\eqref{pp expectation value} that
\al{
\Braket{\bh p^2+m^2}
	&=	{d\ov2\sigma}+\bs P^2+m^2+{d+1\ov2\sigma m^2}\bs P^2+\Or{1\ov\sigma^2},\\
\Braket{\pn{\h p^0}^2}
	&=	-{1\ov2\sigma}+\pn{P^0}^2+{d+1\ov2\sigma m^2}\pn{P^0}^2+\Or{1\ov\sigma^2},
}
and see that there two coincide.

We show the result of the plane-wave expansion with large $\sigma$ in Eq.~\eqref{relativistic covariance}:
\al{
\Braket{\h p^\mu\h p^\nu}-\Braket{\h p^\mu}\Braket{\h p^\nu}
	&=	{1\ov2\sigma}\pn{\eta^{\mu\nu}+{P^\mu P^\nu\ov m^2}}
		+{d\ov8\sigma^2m^2}\pn{\eta^{\mu\nu}+{2P^\mu P^\nu\ov m^2}}
		+\Or{1\ov\sigma^3}.
}
That is,
\al{
\Braket{\h p_i\h p_j}-\Braket{\h p_i}\Braket{\h p_j}
	&=
		{1\ov2\sigma}\pn{\delta_{ij}+{P_iP_j\ov m^2}}
		+{d\ov8\sigma^2m^2}\pn{1+{2P_iP_j\ov m^2}}
		+\Or{1\ov\sigma^3},
			\label{momentum uncertainty simplified}\\
\Braket{\pn{\h p^0}^2}-\Braket{\h p^0}^2
	&=	{\bs P^2\ov2\sigma m^2}
		+{d\ov8\sigma^2m^2}\pn{1+{2\bs P^2\ov m^2}}
		+\Or{1\ov\sigma^3}.
			\label{energy dispersion}
}
If we instead perform the non-relativistic expansion for large $m$ in Eq.~\eqref{relativistic covariance}, we obtain
\al{
\Braket{\h p^i\h p^j}-\Braket{\h p^i}\Braket{\h p^j}
	&=	{\delta^{ij}\ov2\sigma}
			+{P^iP^j\ov2\sigma m^2}
			+{d\ov8\sigma^2m^2}\delta^{ij}
			+\Or{1\ov m^4},
			\label{momentum dispersion}\\
\Braket{\paren{\h p^0}^2}-\Braket{\h p^0}^2
	&=	{\bs P^2\ov2\sigma m^2}
			+{d\ov8\sigma^2m^2}
			+\Or{1\ov m^4}.
}

Several comments are in order:
The first term in Eq.~\eqref{momentum dispersion} reproduces the momentum variance for the ordinary Gaussian wave packet, which is spherically symmetric $\propto\delta_{ij}$. The second term shows that even the off-diagonal covariance for $i\neq j$ is non-zero, due to the boost in the momentum space mentioned above.
The first term in the energy variance~\eqref{energy dispersion} is also due to the boost, and is canceled out when we take the Lorentz invariant combination $\Braket{\h p^2}-\Braket{\h p}^2$ ($=-m^2-\Braket{\h p}^2$): By subtracting both-hand sides of Eq.~\eqref{energy dispersion} from those of Eq.~\eqref{momentum dispersion} contracted with $\delta_{ij}$, we obtain
\al{
-\Braket{\h p}^2
	&=	m^2+{d\ov2\sigma}+{d\paren{d-1}\ov8\sigma^2m^2}+\cdots.
}
As mentioned above, we see that the mass constructed from the expectation value of $\Braket{\h p^\mu}$ is increased from the intrinsic mass $m$.

\subsection{Position (co)variance}
Now let us compute the expectation value $\Braket{\h x^i}$ and its covariance:
\al{
\Braket{
	\Pn{\h x^i-\Braket{\h x^i}}
	\Pn{\h x^j-\Braket{\h x^j}}
	}
	&=	\Braket{\h x^i\h x^j}-\Braket{\h x^i}\Braket{\h x^j}.
}

First,
\al{
\Braket{\h x^i}
	&=	\int{\df^d\bs p\ov2E_{\bs p}}\dBraket{\Braket{X,\bs P|\bs p}}\dBra{\Bra{\bs p}}\h x^i\dKet{\Ket{X,\bs P}}\nn
	&=	\int{\df^d\bs p\ov2E_{\bs p}}\dBraket{\Braket{X,\bs P|\bs p}}i\paren{
		{\p\ov\p p_i}
		-{p^i\ov2E_{\bs p}^2}
		}\dBraket{\Braket{\bs p|X,\bs P}}\nn
	&=	X^i-\Braket{u^i\ov u^0}X^0,
		\label{x expectation value}
}
where we write
\al{
\Braket{u^{i_1}\dots u^{i_\ell}\ov\pn{u^0}^n}
	&:=	m^{n-\ell}\Braket{\h p^{i_1}\dots\h p^{i_\ell}\ov\pn{\h p^0}^n},
}
and we have used 
the following identity:\footnote{
This may be derived for a general (timelike) $D$-vector $\Xi$ as
\als{
\int{\df^d\bs u\ov2u^0}e^{u\cdot\Xi}\paren{
	\Xi^i-\Xi^0{u^i\ov u^0}
	-{u^i\ov\paren{u^0}^2}
	}
	&=	\int{\df^d\bs u\ov2u^0}\paren{
	{\p\ov\p u^i}
	-{u^i\ov\paren{u^0}^2}
	}
	e^{u\cdot\Xi}
	=	{1\ov2}\int\df^d\bs u
		{\p\ov\p u^i}\paren{
			{e^{u\cdot\Xi}\ov u^0}}
	=	0,
}
and then putting the ``on-shell'' value $\Xi=2\sigma mP$.
Recall that $p$ and $u$ are on-shell.
}
\al{
2m\sigma P^i-2m\sigma E_{\bs P}\Braket{u^i\ov u^0}-\Braket{u^i\ov\pn{u^0}^2}
	&=	0.
	\label{first identity}
}

We have defined $\bh x$ as a time-independent Schr\"odinger-picture operator in a certain frame.
Therefore the expectation value~\eqref{x expectation value} should correspond to measurement in an equal-time slice in this frame, and hence the appearance of the non-covariant velocity $\Braket{\bs u\ov u^0}$ rather than the covariant one $\Braket{\bs u\ov m}$.
As we identify the Schr\"odinger, Heisenberg, and interaction pictures at $x^0=0$, the expectation value~\eqref{x expectation value} corresponds to the measurement on the spacelike hyperplane $\Sigma_{(0)}$.
If we instead consider the time-dependent operator $\bh x_\tx{I}\fn{x^0}:=e^{i\h H_\tx{free}x^0}\bh x\,e^{-i\h H_\tx{free}x^0}$ in the interaction picture, we obtain
\al{
\Braket{\h x^i_\tx{I}\fn{x^0}}
	&=	\int{\df^d\bs p\ov2E_{\bs p}}\dBraket{\Braket{X,\bs P|\bs p}}\dBra{\Bra{\bs p}}e^{i\h H_\tx{free}x^0}\h x^ie^{-i\h H_\tx{free}x^0}\dKet{\Ket{X,\bs P}}\nn
	&=	X^i+\Braket{u^i\ov u^0}\pn{x^0-X^0}.
}

Second, we may similarly compute
\al{
\Braket{\h x^i\h x^j}
	&=	\int\df^d\bs p\paren{{\p\ov\p p^i}{\dBK{X,\bs P|\bs p}\ov\sqrt{2E_{\bs p}}}}
		\paren{{\p\ov\p p^j}{\dBK{\bs p|X,\bs P}\ov\sqrt{2E_{\bs p}}}}\nn
	&=	N_\sigma^2m^{d-3}\int\df^d\bs u\paren{{\p\ov\p u^i}{e^{mu\cdot\paren{\sigma P+iX}}\ov\sqrt{2u^0}}}
		\paren{{\p\ov\p u^j}{e^{mu\cdot\paren{\sigma P-iX}}\ov\sqrt{2u^0}}}.
}
Using
$
{\p\ov\p u^i}{e^{mu\cdot\paren{\sigma P+iX}}\ov\sqrt{2u^0}}
	=	e^{imu\cdot X}\sqbr{
			{\p\ov\p u^i}{e^{m\sigma u\cdot P}\ov\sqrt{2u^0}}
			+im{\p\paren{u\cdot X}\ov\p u^i}
				{e^{m\sigma u\cdot P}\ov\sqrt{2u^0}}
			}$,
we can show that $\Im\Braket{\h x^i\h x^j}=0$.
Then we obtain
\al{
\Braket{\h x^i\h x^j}
	&=	\Braket{
				\paren{\sigma\paren{P_i-P^0{u_i\ov u^0}}-{u_i\ov2m\paren{u^0}^2}}
				\paren{\sigma\paren{P_j-P^0{u_j\ov u^0}}-{u_j\ov2m\paren{u^0}^2}}
			+
			\paren{X_i-X^0{u_i\ov u^0}}
			\paren{X_j-X^0{u_j\ov u^0}}
		}\nn
	&=	X_iX_j+\paren{X^0}^2\Braket{u_iu_j\ov\paren{u^0}^2}
		-X_iX^0\Braket{u_j\ov u^0}-X_jX^0\Braket{u_i\ov u^0}\nn
	&\quad
		+\sigma^2P_iP_j
		-\sigma^2\Braket{u_i\ov u^0}P^0P_j
		-\sigma^2\Braket{u_j\ov u^0}P^0P_i
		+\sigma^2\paren{P^0}^2\Braket{u_iu_j\ov\paren{u^0}^2}\nn
	&\quad
		-{\sigma P_i\ov2m}\Braket{u_j\ov\paren{u^0}^2}
		-{\sigma P_j\ov2m}\Braket{u_i\ov\paren{u^0}^2}
		+{\sigma P^0\ov m}\Braket{u_iu_j\ov\paren{u^0}^3}
		+{1\ov4m^2}\Braket{u_iu_j\ov\paren{u^0}^4},
}
and hence\footnote{
The last three terms in Eq.~\eqref{x co-variance} may be recast into the form ${1\ov4m^2}\Braket{u_iu_j\pn{1+\Xi^0u^0}^2\ov\paren{u^0}^4}$ with $\Xi=2\sigma mP$ but we compute it as is.
}
\al{
\Braket{\h x^i\h x^j}
	-\Braket{\h x^i}\Braket{\h x^j}
	&=	\paren{X^0}^2\paren{\Braket{u_iu_j\ov\paren{u^0}^2}-\Braket{u_i\ov u^0}\Braket{u_j\ov u^0}}\nn
	&\quad
		-\sigma^2P_iP_j
		+\sigma^2\paren{P^0}^2\Braket{u_iu_j\ov\paren{u^0}^2}
		+{\sigma P^0\ov m}\Braket{u_iu_j\ov\paren{u^0}^3}
		+{1\ov4m^2}\Braket{u_iu_j\ov\paren{u^0}^4},
			\label{x co-variance}
}
where we have used the identities~\eqref{first identity} and 
\al{
0	&=	2m\sigma P_i\Braket{1\ov\paren{u^0}^{n-1}}
		-2m\sigma P^0\Braket{u_i\ov\paren{u^0}^n}
		-n\Braket{u_i\ov\paren{u^0}^{n+1}}.
}

In this paper, we compute the expectation value~\eqref{x expectation value} and the (co)variance~\eqref{x co-variance} using the saddle-point method for large $\sigma$:
\al{
\Braket{\h x^i}
	&=	X^i-X^0{P^i\ov P^0}\sqbr{1-{1\ov2\sigma\pn{P^0}^2}+\Or{1\ov\sigma^2}},\\
\Braket{\h x^i\h x^j}
	-\Braket{\h x^i}\Braket{\h x^j}
	&=	{\sigma\ov2}\pn{\delta_{ij}-{P_iP_j\ov\pn{P^0}^2}}+\Or{\sigma^0}.
		\label{position uncertainty final result}
}
Especially for the variance $i=j$,
\al{
\Braket{\pn{\h x_i}^2}-\Braket{\h x_i}^2
	&=	{\sigma\ov2}\pn{1-{P_i^2\ov E_{\bs P}^2}}+\Or{\sigma^0},
		\label{x uncertainty}
}
where $i$ is not summed.
One may find the detailed derivation in Appendix~\ref{basic integral}.
Especially, we have used Eq.~\eqref{saddle-point 1st} to compute
$\Braket{u^i\ov u^0}={\mc I_{(1)}^i\ov\mc I_{(0)}}={\Xi^i\ov\Xi^0}\paren{
			1-{\abb\Xi\ov\pn{\Xi^0}^2}+\cdots}$ with $\Xi=2m\sigma P$ and $\abb{\Xi}=2\sigma m^2$ for $\Braket{\h x^i}$, and similarly Eq.~\eqref{saddle-point 2nd} for $\Braket{\h x^i\h x^j}$.
The result~\eqref{x uncertainty} implies that if we measure the position uncertainty along the direction of $\bs P$, it is Lorentz-contracted by the factor $m/E_{\bs P}$, compared to the measurement transverse to $\bs P$~\cite{Naumov:2010um,Naumov:2013uia,Naumov:2020yyv}.

\subsection{Uncertainty relation}
Finally combining Eqs.~\eqref{momentum uncertainty simplified} and \eqref{x uncertainty}, the uncertainty relation on the time slice $\Sigma_{(0)}$ becomes
\al{
\sqrt{\Braket{\pn{\h x_i}^2}-\Braket{\h x_i}^2}
\sqrt{\Braket{\pn{\h p_i}^2}-\Braket{\h p_i}^2}
	&=	{1\ov2}\sqrt{\pn{1-{P_i^2\ov E_{\bs P}^2}}\pn{1+{P_i^2\ov m^2}}}
		+\Or{1\ov\sigma},
}
where $i$ is not summed.
In the non-relativistic limit, we see that the terms of order $\bs P^2/m^2$ cancel out:
\al{
\sqrt{\Braket{\pn{\h x_i}^2}-\Braket{\h x_i}^2}
\sqrt{\Braket{\pn{\h p_i}^2}-\Braket{\h p_i}^2}
	&=	{1\ov2}+\Or{\pn{\bs P^2}^2\ov m^4},
}
where $i$ is not summed.
The ordinary minimum uncertainty for the Gaussian wave is recovered in the non-relativistic limit.

When we measure along a direction $\bs n$ with $\ab{\bs n}=1$ and $\bs n\cdot\bs P=\ab{\bs P}\cos\theta$,
\al{
\sqrt{\Braket{\pn{\h x_{\bs n}}^2}-\Braket{\h x_{\bs n}}^2}
\sqrt{\Braket{\pn{\h p_{\bs n}}^2}-\Braket{\h p_{\bs n}}^2}
	&=	{1\ov2}\sqrt{\pn{1-{\ab{\bs P}^2\cos^2\theta\ov E_{\bs P}^2}}\pn{1+{\ab{\bs P}^2\cos^2\theta\ov m^2}}}
		+\Or{1\ov\sigma},
}
where we write $A_{\bs n}:=\bs n\cdot\bs A$ for any spatial vector $\bs A$.
We see that the uncertainty is minimized to $1/2$ when we measure along the directions $\theta=0$, $\pi/2$, and $\pi$; namely, either when it is parallel or perpendicular to $\bs P$.


\subsection{Time-energy uncertainty}
Before proceeding, we comment on the time uncertainty. The Lorentz-invariant wave packet is not localized in time as discussed above, and therefore the expectation value of time ``$\h x^0$'' for this wave packet is ill-defined, just as the expectation value of position is ill-defined for a plane wave.\footnote{
In the current on-shell formulation, the free one-particle Hilbert space is spanned within a $d$-dimensional spatial hyperplane that is a fixed-time surface in a certain Lorentz frame: On an arbitrary spatial hyperplane $\Sigma$, integral of the probability density becomes unity, $\int_\Sigma\df\Sigma^\mu\dBK{X,\bs P|x}{2i\p_\mu}\dBK{x|X,\bs P}=1$; see Eq.~\eqref{completeness, Lorentz invariant}.
It might be interesting to develop an off-shell formulation spanned within the whole enlarged $D$-dimensional spacetime and to discuss the uncertainty relation between time and energy more directly there.
}
However, we can show that the energy uncertainty~\eqref{energy dispersion} is matched with an uncertainty of the time at which this wave packet passes through a certain point $\bs x$.
Suppose we are at $\bs x=\bs X$ and see the wave packet passing through it around the time $x^0\sim X^0$. Then the probability density on each $\Sigma_{(x^0)}$ along the worldline $\bs x=\bs X$ becomes
\al{
\ab{\dB{x}\sqrt{2E_{\bh p}}\dK{X,\bs P}}^2
	&\simeq
		{\paren{\sigma\ov\pi}^{d\ov2}E_{\bs P}
			\ov\paren{\sigma m}^{2d-1}}
		e^{-{\bs P^2\ov\sigma m^2}\pn{x^0-X^0}^2}.
}
From the exponent, we see that the timelike width-squared is $\pn{\Delta t}^2\simeq{\sigma m^2\ov2\bs P^2}$.
Comparing with the energy uncertainty~\eqref{energy dispersion}, we see that the time-energy uncertainty takes the minimum value for the position-momentum one at the leading order:
\al{
\Delta t\,\Delta p^0
	&\simeq	{1\ov2}.
}

It would also be interesting to consider a wave-packet scattering. Then what is localized in time is not each wave packet but an \emph{overlap} of the wave packets: This kind of timelike width-squared of the overlap region is given as $\sigma_t$ in Ref.~\cite{Ishikawa:2018koj} and as $\varsigma_\tx{in},\varsigma_\tx{out}$ in Refs.~\cite{Ishikawa:2020hph,Ishikawa:2021bzf} for the Gaussian wave packets. This will be pursued in a separate publication.

\section{Completeness of Lorentz-invariant basis}\label{completeness section}

Let us discuss the completeness on $L^2\fn{\mathbb R^d}$.
We will prove the following manifestly Lorentz-invariant completeness relation:
\al{
\int_{\Sigma_X}{\df^d\Sigma_X^\mu\ov\pn{2\pi}^d}\int{\df^d\bs P\ov2E_{\bs P}}
	\dBK{\bs p|X,\bs P}\pn{2i{\p\ov\p X^\mu}}\dBK{X,\bs P|\bs q}
	&=	\dBK{\bs p|\bs q},
		\label{invariant completeness}
}
where
$\Sigma_X$ is an arbitrary spacelike hyperplane in the $X$ space and
$\df^d\Sigma_X^\mu$ is a $d$-volume element that is normal to $\Sigma_X$.
In the language of differential forms, $\df^d\Sigma_{X\mu}=-\,{\star\df X_\mu}=-{1\ov d!}\epsilon_{\mu\mu_1\dots\mu_d}\df X^{\mu_1}\wedge\cdots\wedge\df X^{\mu_d}$; 
see Eq.~\eqref{completeness, Lorentz invariant} and below; 
see also Ref.~\cite{Debbasch:2007rs} for discussion on Lorentz invariance of the phase space volume.
In other words,
\al{
\int_{\Sigma_X}{\df^d\Sigma_X^\mu\ov\pn{2\pi}^d}\int{\df^d\bs P\ov2E_{\bs P}}
	\dK{X,\bs P}\pn{2i{\p\ov\p X^\mu}}\dB{X,\bs P}
	&=	\h 1,
		\label{Lorentz-covariant completeness}
}
where the right-hand side is the identity operator in the free one-particle subspace.
Now the Lorentz-friendly plane wave is expanded as
\al{
\dK{\bs p}
	&=	\int_{\Sigma_X}{\df^d\Sigma_X^\mu\ov\pn{2\pi}^d}\int{\df^d\bs P\ov2E_{\bs P}}
	\dK{X,\bs P}\pn{2i{\p\ov\p X^\mu}}\dBK{X,\bs P|\bs p}.
}

The completeness~\eqref{invariant completeness} can be rewritten as
\al{
N_\sigma^2\int_{\Sigma_X}{-2q_\mu\df^d\Sigma_X^\mu\ov\paren{2\pi}^d}
\int{\df^d\bs P\ov 2E_{\bs P}}
	e^{-ip\cdot\pn{X+i\sigma P}}
	e^{iq\cdot\pn{X-i\sigma P}}
	&=	2E_{\bs p}\delta^d\fn{\bs p-\bs q},
	\label{Lorentz invariant completeness}
}
The proof of Eq.~\eqref{Lorentz invariant completeness} is as follows:
Noting that the left-hand side of Eq.~\eqref{Lorentz invariant completeness} is manifestly Lorentz invariant (recall Eq.~\eqref{phi expanded}), we may choose $\Sigma$ to be a constant $X^0$-plane without loss of generality:
\al{
N_\sigma^2
\int{2E_{\bs q}\df^d\bs X\ov\paren{2\pi}^d}{\df^d\bs P\ov 2E_{\bs P}}
	e^{i\paren{E_{\bs p}-E_{\bs q}}X^0}
	e^{-i\paren{\bs p-\bs q}\cdot\bs X}
	e^{\sigma\paren{p+q}\cdot P}
&=	2E_{\bs p}\delta^d\fn{\bs p-\bs q}
	N_\sigma^2
	\int{\df^d\bs P\ov 2E_{\bs P}}
	e^{2\sigma p\cdot P}\nn
&=	2E_{\bs p}\delta^d\fn{\bs p-\bs q}.
}

We may rewrite the completeness relation~\eqref{Lorentz-covariant completeness} in a different fashion:
\al{
M_\sigma^2\int_{\Sigma_X}{\df^d\Sigma_X^\mu\ov\pn{2\pi}^d}
\int{\df^d\bs P\ov2E_{\bs P}}\pn{-2P_\mu}
	\dK{X,\bs P}\dB{X,\bs P}
	&=	\h 1,
}
where
\al{
M_\sigma
	&:=	\sqrt{K_{d-1\ov2}\fn{2\sigma m^2}\ov K_{d+1\ov2}\fn{2\sigma m^2}}.
}
To show this, we may sandwich both-hand sides by $\dB{\bs p}$ and $\dK{\bs q}$, take the frame where $\Sigma_X$ becomes a constant-$X^0$ plane, and use Eq.~\eqref{I mu}.\footnote{
This computation is inspired by Ref.~\cite{Kaiser:1977ys}.
}

The completeness~\eqref{Lorentz-covariant completeness} on the one-particle subspace $L^2\fn{\mathbb R^d}$ can be naturally generalized to that on the whole Fock space $\mc H$ as follows.
When we define $\cA_{X,\bs P}$ by
\al{
\cA_{X,\bs P}^\dagger\Ket{0}
	&:=	\dKet{\Ket{X,\bs P}},
}
with mass dimensions $\sqbr{\cA_{X,\bs P}}=\Sqbr{\dKet{\Ket{X,\bs P}}}=0$,
we obtain\footnote{
To show it, we first expand as
$
\cA_{X,\bs P}
	=	\int{\df^d\bs p\ov2E_{\bs p}}f_{\bs p}\fn{X,\bs P}\wh\alpha_{\bs p}
$.
Putting this into
$
\dBra{\Bra{\bs p}}\cA_{X,\bs P}^\dagger\Ket{0}
	=	\dBraket{\Braket{\bs p|X,\bs P}}
$,
we get
$
f_{\bs p}^*\fn{X,\bs P}
	=	\dBraket{\Braket{\bs p|X,\bs P}}
$,
that is,
$
f_{\bs p}\fn{X,\bs P}
	=	\dBraket{\Braket{X,\bs P|\bs p}}
$.
}
\al{
\cA_{X,\bs P}
	&=	\int{\df^d\bs p\ov2E_{\bs p}}\dBraket{\Braket{X,\bs P|\bs p}}\wh\alpha_{\bs p},
			\label{expansion of A}
}
and
\al{
\wh\alpha_{\bs p}
	&=	\int_\Sigma{\df^d\Sigma^\mu\ov\pn{2\pi}^d}{\df^d\bs P\ov2E_{\bs P}}
		\pn{-2i{\p\ov\p X^\mu}\dBK{\bs p|X,\bs P}}
		\cA_{X,\bs P}.
		\label{alpha in terms of A}
}
Now we get
\al{
\commutator{\cA_{X,\bs P}}{\cA_{X',\bs P'}^\dagger}
	&=	\int{\df^d\bs p\ov2E_{\bs p}}\dBraket{\Braket{X,\bs P|\bs p}}
		\int{\df^d\bs p'\ov2E_{\bs p'}}\dBraket{\Braket{\bs p'|X',\bs P'}}
		\commutator{\wh\alpha_{\bs p}}{\wh\alpha_{\bs p'}^\dagger}\nn
	&=	\int{\df^d\bs p\ov2E_{\bs p}}\dBraket{\Braket{X,\bs P|\bs p}}
		\dBraket{\Braket{\bs p|X',\bs P'}}\wh1\nn
	&=	\dBraket{\Braket{X,\bs P|X',\bs P'}}\wh1.
}

Putting Eq.~\eqref{alpha in terms of A} into the expansion~\eqref{phi expanded}, we obtain
\al{
\wh\phi\fn{x}
	&=	\int{\df^d\bs p\ov2E_{\bs p}}\sqbr{
			\dBraket{\Braket{x|\bs p}}
			\int_\Sigma{\df^d\Sigma^\mu\ov\pn{2\pi}^d}{\df^d\bs P\ov2E_{\bs P}}
		\pn{-2i{\p\ov\p X^\mu}\dBK{\bs p|X,\bs P}}
		\cA_{X,\bs P}
			+\tx{h.c.}
			}\nn
	&=	\sqbr{
			\int_\Sigma{\df^d\Sigma^\mu\ov\pn{2\pi}^d}{\df^d\bs P\ov2E_{\bs P}}
		\pn{-2i{\p\ov\p X^\mu}\dBK{x|X,\bs P}}
		\cA_{X,\bs P}
			+\tx{h.c.}
			}.
}
Using
$
{\df\ov\df z}{K_{d-1\ov2}\fn{z}\ov z^{d-1\ov2}}
	=	-{K_{d+1\ov2}\fn{z}\ov z^{d-1\ov2}}$,
we may obtain the explicit form of the above expansion coefficient:
\al{
{\p\ov\p X^\mu}\dBK{x|X,\bs P}
	&=	
		{1\ov\paren{4\pi\sigma}^{d-1\ov4}}
		{\paren{2\sigma m^2}^{d-1\ov2}
			\ov\sqrt{2\pi K_{d-1\ov2}\fn{2\sigma m^2}}}
		{K_{d+1\ov2}\fn{\abb\Xi}\ov
		\abb\Xi^{{d+1\ov2}}}
		m\pn{x-X-i\sigma P}_\mu,
}
where $\abb\Xi=m\sqrt{\pn{x-X-i\sigma P}^2}$; see Eqs.~\eqref{Xi for wave function} and \eqref{argument Xi}.

The branch cut for the square-root in the argument is along
\al{
\paren{x-X}\cdot P
	&=	0,	&
\paren{x-X}^2-\sigma^2P^2
	&<0,
}
that is,
\al{
\paren{x-X}\cdot P
	&=	0,	&
\paren{\bs x-\bs X}^2+\sigma^2m^2
	&<\paren{x^0-X^0}^2.
}
In a coordinate system $x'$ that is a rest frame for $P$, the cut is along
\al{
x^{\pr0}
	&=	X^{\pr0},	&
\paren{\bs x'-\bs X'}^2+\sigma^2m^2
	&<	0.
}
This is never satisfied and hence we are never on the cut; see also the last paragraph in Appendix~\ref{Master integral section}.

To cultivate some intuition, we show the case of $\Sigma_X$ being a constant-$X^0$ plane:
\al{
\wh\phi\fn{x}
	&=	
		{2m^d\pn{\sigma\ov\pi}^{d-1\ov4}
			\ov\sqrt{2\pi K_{d-1\ov2}\fn{2\sigma m^2}}}
		\sqbr{
			\int_{\Sigma_X}{\df^d\bs X\ov\pn{2\pi}^d}{\df^d\bs P\ov2E_{\bs P}}
		{K_{d+1\ov2}\fn{\abb\Xi}\ov
		\abb\Xi^{{d+1\ov2}}}
		\pn{\sigma E_{\bs P}+i\pn{x^0-X^0}}
		\cA_{X,\bs P}
			+\tx{h.c.}
			}.
}

\section{Summary and discussion}
We have proposed a Lorentz-invariant generalization of the Gaussian wave packet. This Lorentz-invariant wave packet has more natural dependence on the central position $X$ and momentum $P$ than the coherent state in the position-momentum space: The dependence is holomorphic through the variable $X+i\sigma P$ if we further generalize it for a time-like off-shell momentum $P$.

We have obtained the wave function for the Lorentz-invariant wave packet in an closed analytic form, as well as their inner products. The wave function is localized in space but not in time, while its width becomes larger and larger as the time is more and more apart from $X^0$.

We have computed the expectation value and (co)variance of momentum for this state in a closed analytic form, and those of position in the saddle-point approximation.
They reduce to the minimum position-momentum uncertainty of the corresponding Gaussian wave packet in the non-relativistic limit.
The time-energy uncertainty takes the same minimum value at the leading order in the large width expansion.

We have managed to obtain the completeness relation for these Lorentz-invariant wave packets in a manifestly Lorentz-invariant fashion.
It would be interesting to use the complete set of the Lorentz-invariant wave packets instead of the Gaussian ones in the decay and scattering processes analyzed so far.
It would be worth applying to the wave packets in neutrino physics too.

\subsection*{Acknowledgement}
We thank Koichi Hamaguchi, Kenzo Ishikawa, and Kenji Nishiwaki for useful discussion. The work of K.O.\ is in part supported by JSPS Kakenhi Grant No.~19H01899.

\appendix
\section*{Appendix}

\section{More on coherent states}\label{coherent appendix}
We define the displacement operator:
\al{
\h D\fn{\bs\alpha}
	&:=	e^{\bs\alpha\cdot\bh a^\dagger-\bs\alpha^*\cdot\bh a\ov2\ab{\lambda}^2}.
}
Note that $\h D^\dagger\fn{\bs\alpha}=\h D\fn{-\bs\alpha}$.
Using the Baker-Campbell-Hausdorff formula 
\al{
e^{\h A}e^{\h B}
	&=	e^{\h A+\h B+{1\ov2}\Commutator{\h A}{\h B}}
}
for the case of $\Commutator{\h A}{\h B}$ being a (commuting) number, we can show that $\h D$ is unitary, $\h D\h D^\dagger=\h D^\dagger\h D=\h1$, and that
\al{
\h D\fn{\bs\alpha}
	&=	e^{-{\ab{\bs\alpha}^2\ov4\ab{\lambda}^2}}
		e^{\bs\alpha\cdot\bh a^\dagger\ov2\ab{\lambda}^2}
		e^{-{\bs\alpha^*\cdot\bh a\ov2\ab{\lambda}^2}}
	=	e^{\ab{\bs\alpha}^2\ov4\ab{\lambda}^2}
		e^{-{\bs\alpha^*\cdot\bh a\ov2\ab{\lambda}^2}}
		e^{\bs\alpha\cdot\bh a^\dagger\ov2\ab{\lambda}^2},\\
\h D\fn{\bs\alpha}\h D\fn{\bs\beta}
	&=	e^{\bs\alpha\cdot\bs\beta^*-\bs\alpha^*\cdot\bs\beta\ov4\ab{\lambda}^2}\h D\fn{\bs\alpha+\bs\beta},
}
where we have defined $\ab{\bs\alpha}^2:=\bs\alpha^*\cdot\bs\alpha=\sum_{i=1}^d\ab{\alpha_i}^2$.
We then get
\al{
\Braket{\bs\alpha|\bs\beta}
	&=	\Bra{\varphi}\h D\fn{-\bs\alpha}\h D\fn{\bs\beta}\Ket{\varphi}
	=	e^{-\bs\alpha\cdot\bs\beta^*+\bs\alpha^*\cdot\bs\beta\ov4\ab{\lambda}^2}
		\Bra{\varphi}\h D\fn{-\bs\alpha+\bs\beta}\Ket{\varphi}
	=	e^{-\bs\alpha\cdot\bs\beta^*+\bs\alpha^*\cdot\bs\beta\ov2}
		e^{-{\ab{\bs\alpha-\bs\beta}^2\ov2}}\nn
	&=	e^{-{\ab{\bs\alpha}^2+\ab{\bs\beta}^2\ov4\ab{\lambda}^2}}e^{\bs\alpha^*\cdot\bs\beta\ov2\ab{\lambda}^2}.
}

The ground state of the harmonic oscillator $\Ket{\varphi}$ is given by
\al{
\bh a\Ket{\varphi}
	&=	0.
}
That is, $\Ket{\varphi}=\bigl.\Ket{\bs\alpha}\bigr|_{\bs\alpha=\bs0}$, namely
\al{
\Braket{\bs x|\varphi}
	&=	{1\ov\paren{\pi\sigma}^{d\ov4}}e^{-{\bs x^2\ov2\sigma}},&
\Braket{\varphi|\bs p}
	&=	\paren{\pi\ov\sigma}^{d\ov4}e^{-{\sigma\ov2}\bs p^2}.
}
From the commutator 
\al{
\commutator{\wh a_i}{\h D\fn{\bs\alpha}}
	&=	\sum_{n=0}^\infty{1\ov n!}\commutator{\wh a_i}{\paren{\bs\alpha\cdot\bh a^\dagger-\bs\alpha\cdot\bh a\ov2\ab{\lambda}^2}^n}
	=	\sum_{n=1}^\infty{1\ov n!}n{\alpha_j\Commutator{\wh a_i}{\wh a^\dagger_j}\ov2\ab{\lambda}^2}\paren{\bs\alpha\cdot\bh a^\dagger-\bs\alpha\cdot\bh a\ov2\ab{\lambda}^2}^{n-1}\nn
	&=	\alpha_i\h D\fn{\bs\alpha},
}
we see that $\h a_i\h D\fn{\bs\alpha}\Ket{\varphi}=\alpha_i\h D\fn{\bs\alpha}\Ket{\varphi}$, namely,
\al{
\Ket{\bs\alpha}
	&=	\h D\fn{\bs\alpha}\Ket{\varphi}
}
and that the similarity transformation of the annihilation operator results in its displacement:
\al{
\h D^\dagger\fn{\bs\alpha}\bh a\h D\fn{\bs\alpha}
	&=	\bh a+\bs\alpha.
}
In this sense, $\Ket{\bs\alpha}$ is the ground state of the harmonic oscillator displaced by $\bs\alpha$.

\section{Master integral}\label{basic integral}
We encounter an integral of the form
\al{
I\fn{\check\Xi}
	&=	\int{\df^d\bs p\ov2E_{\bs p}}e^{p\cdot\check\Xi}
}
where $p$ is on-shell and $\check\Xi$ is an arbitrary complex $D$-vector of mass dimension $-1$.
For a massive particle $p^2=-m^2<0$, it is more convenient to use the $D$-velocity $u:=p/m$ as an integration variable. Hereafter, we always take $u$ ``on-shell'', $u^2=-1$ and $u^0=\sqrt{1+\bs u^2}$, unless otherwise stated. 
Let us define
\al{
\mc I\fn{\Xi}
	&:=	\int{\df^d\bs u\ov2u^0}e^{u\cdot\Xi}
	=	\int\df^Du\,\delta\fn{u^2+1}\theta\fn{u^0}e^{u\cdot\Xi},
}
where $u$ is on-shell and off-shell for the first and second integrals, respectively, and
$\Xi$ is a dimensionless complex $D$-vector.
Trivially substituting $\Xi=m\check\Xi$, we get
\al{
I\fn{\check\Xi}
	&=	m^{d-1}\mc I\fn{m\check\Xi}.
}
So far we have not put any kind of on-shell condition on $\Xi$, and hence
\al{
{\p\abb{\Xi}\ov\p\Xi^\mu}
	&=	-{\Xi_\mu\ov\abb{\Xi}},
}
where $\abb\Xi:=\sqrt{-\Xi^2}=\sqrt{\pn{\Xi^0}^2-\bs\Xi^2}$ as given in the main text.

We write $\Re\Xi=:\mc P$ and $\Im\Xi=:\mc Q$, which later will correspond to some momentum and position, respectively:
\al{
\Xi=\mc P+i\mc Q.
	\label{Xi P Q}
}
Trivially, $\Xi^2=\paren{\mc P+i\mc Q}^2=\mc P^2+2i\mc P\cdot\mc Q-\mc Q^2$.

\subsection{Evaluation of master integral}\label{Master integral section}
We focus on the case of $\mc P$ being timelike, $\mc P^2=-\paren{\mc P^0}^2+\bs{\mc P}^2<0$, and future-oriented, $\mc P^0>0$, so that there exists a (proper orthochronous) Lorentz transformation $\Lambda$ to the ``rest frame'' $\wt{\mc P}=\Paren{\wt{\mc P}^0,\bs 0}$ with $\wt{\mc P}^0=\abb{\mc P}>0$ such that $\Lambda\mc P=\wt{\mc P}$. 
For a given $\mc P$, concrete form of $\Lambda$ is, in the matrix notation,
\al{
\Lambda
	&=	\bmat{
			U^0&-\bs U^\t\\
			-\bs U&\wh 1+\paren{U^0-1}\bh U\bh U^\t
			},
}
where $U:=\mc P/\abb{\mc P}$,
$\bh U:=\bs U/\ab{\bs U}$,
$\wh1$ is the $d$-dimensional unit matrix, and
$U^0=\sqrt{1+\bs U^2}$;
see e.g.\ Ref.~\cite{Sogebu}.
We write $\wt{\mc Q}:=\Lambda\mc Q$ and $\wt\Xi:=\wt{\mc P}+i\wt{\mc Q}$. Note that $\bs{\wt{\mc P}}=0$ implies $\bs{\wt\Xi}=i\bs{\wt{\mc Q}}$.
On the other hand, we leave $\mc Q$ to be an arbitrary real $D$ vector.

We change the integration variable to $\wt u:=\Lambda u$. Using the Lorentz invariance of the integration measure etc.\ as well as $\paren{\Lambda^{-1}\wt u}\cdot\Xi=\wt u\cdot\paren{\Lambda\Xi}=\wt u\cdot\wt\Xi$, we get
\al{
\mc I
	&=	\int\df^D\wt u\,\delta\fn{\wt u^2+1}\theta\fn{\wt u^0}e^{\wt u\cdot\wt\Xi}.
}
Using $\wt u\cdot\wt\Xi=-\wt u^0\wt\Xi^0+\wt{\bs u}\cdot\bs{\wt\Xi}=-\wt u^0\wt\Xi^0+i\wt{\bs u}\cdot\bs{\wt{\mc Q}}$ and renaming $\wt{\bs u}$ by $\bs u$, we get
\al{
\mc I
	&=	\int{\df^d\bs u\ov2\sqrt{1+\bs u^2}}e^{-\sqrt{1+\bs u^2}\wt\Xi^0+i\bs u\cdot\bs{\wt{\mc Q}}}\nn
	&=	\Omega_{d-1}
		\int_0^\pi{\sin^{d-2}\theta\,\df\theta\ov\sqrt{\pi}\Gamma\fn{d-1\ov2}/\Gamma\fn{d\ov2}}
		\int_0^\infty{\tf u^{d-1}\df\tf u\ov 2\sqrt{1+\tf u^2}}
		e^{-\sqrt{1+\tf u^2}\wt\Xi^0+i\tf u\ab{\bs{\wt{\mc Q}}}\cos\theta},
}
where $\Omega_{d-1}=2\pi^{d\ov2}/\Gamma\fn{d\ov2}$ is the area of a unit $\paren{d-1}$-sphere (boundary of unit $d$-ball).

We follow Ref.~\cite{Zhang:2008jy} in the following.
The angular integral reads
\al{
\int_0^\pi\sin^{d-2}\theta\,\df\theta\,e^{i\tf u\ab{\bs{\wt{\mc Q}}}\cos\theta}
	&=	\sqrt\pi\,\Gamma\fn{d-1\ov2}\paren{2\ov\tf u\Ab{\bs{\wt{\mc Q}}}}^{d-2\ov2}J_{d-2\ov2}\fn{\tf u\Ab{\bs{\wt{\mc Q}}}},
}
where $J$ is the Bessel function of the first kind.
(For $d=3$, the right-hand side comes back to the familiar form $2\sin\fn{\tf u\Ab{\bs{\wt{\mc Q}}}}/\tf u\Ab{\bs{\wt{\mc Q}}}$.)
Now
\al{
\mc I
	&=	{\Omega_{d-1}\Gamma\fn{d\ov2}\ov2}
		\paren{2\ov\Ab{\bs{\wt{\mc Q}}}}^{d-2\ov2}
		\int_1^\infty\paren{\vep^2-1}^{d-2\ov2}\df\vep\,
		J_{d-2\ov2}\fn{\sqrt{\vep^2-1}\Ab{\bs{\wt{\mc Q}}}}
		e^{-\vep\wt\Xi^0},
}
where $\vep:=\sqrt{1+\tf u^2}$ is a rescaled energy.
We use the second formula of Eq.~(6.645) in Ref.~\cite{GR_book}:
\al{
\int_1^\infty\df x\paren{x^2-1}^{\nu\ov2}e^{-\alpha x}J_\nu\fn{\beta\sqrt{x^2-1}}
	&=	\sqrt{2\ov\pi}\beta^\nu\paren{\alpha^2+\beta^2}^{-{2\nu+1\ov4}}
		K_{2\nu+1\ov2}\fn{\sqrt{\alpha^2+\beta^2}},
}
applicable for $\Re\alpha>0$.
Setting $\alpha=\wt\Xi^0$, $\beta=\ab{\bs{\wt{\mc Q}}}$, and $\nu={d-2\ov2}$, we get
\al{
\mc I\fn{\Xi}
	&=	\pn{2\pi}^{d-1\ov2}
		{K_{d-1\ov2}\Fn{\abb{\Xi}}\ov
		\abb{\Xi}^{{d-1\ov2}}},
}
where we used
\al{
\Pn{\wt\Xi^0}^2+\bs{\wt{\mc Q}}^2
	&=	\Pn{\wt{\mc P}^0}^2
		+2i\wt{\mc P}^0\wt{\mc Q}^0
		-\Pn{\wt{\mc Q}^0}^2
		+\bs{\wt{\mc Q}}^2
	=	-\wt{\mc P}^2-2i\wt{\mc P}\cdot\wt{\mc Q}+\wt{\mc Q}^2
	=	-\wt\Xi^2
	=	-\Xi^2
	=	\abb{\Xi}^2.
}

The limit $\abb{\Xi}\to\infty$ is
\al{
\mc I
	&\to	
		\pn{2\pi}^{d-1\ov2}
		\sqrt{\pi\ov2}{e^{-\abb{\Xi}}\ov \abb{\Xi}^{d\ov2}}\paren{1+\Or{1\ov \abb{\Xi}}},
}
whereas $\abb{\Xi}\to0$ gives
\al{
\mc I
	&\to
			\pn{2\pi}^{d-1\ov2}
			{2^{d-3\ov2}\Gamma\fn{d-1\ov2}\ov \abb{\Xi}^{d-1}}\Pn{1+\Or{\abb{\Xi}}}
			+\Or{\ln \abb{\Xi}}.
}
Recall that we are assuming $d\geq2$.

Throughout this paper, we choose to place a branch-cut for a square root, say $\sqrt{z}$, on the negative real axis of $z$-plane:
For $-\pi<\theta<\pi$ and $r\geq0$,
\al{
\sqrt{re^{i\theta}}
	&:=	\sqrt{r}e^{i\theta/2}.
}
In particular we may use the following limit for $y\to0$ under $x>0$:
\al{
\sqrt{-x-iy}
	&\to	-i\sgn\fn{y}\sqrt{x}\paren{1+i{y\ov2x}+\cdots}.
}
Then for the expression~\eqref{master formula}, the condition on the argument to be on the real axis is
\al{
0	&\sr{!}{=}
		\Im\paren{-\Xi^2}
	=	\Im\paren{-\wt\Xi^2}
	=	2\wt{\mc P}^0\wt{\mc Q}^0.
}
As $\wt{\mc P}^0>0$, we see that $\wt{\mc Q}^0=0$ is it.
However, the real part of the argument on the real axis ($\wt{\mc Q}^0=0$) is positive:
\al{
\Re\paren{-\Xi^2}
	&=	\Re\paren{-\wt\Xi^2}
	=	\paren{\wt{\mc P}^0}^2+\bs{\wt{\mc Q}}^2
	>	0.
}
To summarize, no ambiguity arises from the branch cut as long as $\mc P$ is timelike: $\mc P^2<0$.
More in general, we may perform analytic continuation of the result~\eqref{master formula} so long as $\Im\paren{-\Xi^2}\propto\mc P\cdot\mc Q\neq0$ or $\Re\paren{-\Xi^2}=\mc Q^2-\mc P^2>0$.
On the other hand, possible non-triviality arises when $\mc Q^2<\mc P^2$ ($<0$) in the limit $\mc P\cdot\mc Q\to0$:
\al{
\sqrt{-\Xi^2}
	&=	\sqrt{-\mc P^2-2i\mc P\cdot\mc Q+\mc Q^2}
	\to	-i\sgn\fn{\mc P\cdot\mc Q}\sqrt{\mc P^2-\mc Q^2}\paren{1+i{\mc P\cdot\mc Q\ov\mc P^2-\mc Q^2}+\cdots}.
}

\subsection{Derivative of master integral}\label{derivative of master integral}
We define
\al{
\mc I^{\mu_1\dots\mu_n}\fn{\Xi}
	&:=	{\p^n\ov\p\Xi_{\mu_1}\cdots\p\Xi_{\mu_n}}\mc I\fn{\Xi}.
}
We note that
\al{
\mc I^{\mu_1\dots\mu_n}\fn{\Xi}
	&=	
		\int{\df^d\bs u\ov2u^0}e^{u\cdot\Xi}u^{\mu_1}\cdots u^{\mu_n}\nn
	&=	
		\int\df^Du\,\delta\fn{u^2+1}\theta\fn{u^0}e^{u\cdot\Xi}
		u^{\mu_1}\cdots u^{\mu_n},
}
where $u$ is on-shell and off-shell in the first and second integrals, respectively.
Using
\al{
{\df\mc I\ov \df\abb{\Xi}}
	&=	-\paren{2\pi}^{d-1\ov2}{K_{d+1\ov2}\fn{\abb{\Xi}}\ov
		\abb{\Xi}^{{d-1\ov2}}},&
{\df^2\mc I\ov \df\abb{\Xi}^2}
	&=	\paren{2\pi}^{d-1\ov2}\sqbr{
		-{K_{d+1\ov2}\fn{\abb{\Xi}}\ov \abb{\Xi}^{d+1\ov2}}
		+{K_{d+3\ov2}\fn{\abb{\Xi}}\ov \abb{\Xi}^{d-1\ov2}}
		},
}
and
\al{
{\p \abb{\Xi}\ov\p\Xi_\mu}
	&=	-{\Xi^\mu\ov \abb{\Xi}},&
{\p^2\abb{\Xi}\ov\p\Xi_\mu\p\Xi_\nu}
	&=	-{1\ov \abb{\Xi}}\paren{\eta^{\mu\nu}-{\Xi^\mu\Xi^\nu\ov\Xi^2}},
}
we obtain
\al{
\mc I^\mu
	=	{\p\mc I\ov\p\Xi_\mu}
	&=	\paren{2\pi}^{d-1\ov2}
		{K_{d+1\ov2}\fn{\abb{\Xi}}\ov
		\abb{\Xi}^{{d+1\ov2}}}
		\Xi^\mu,
		\label{I mu}\\
\mc I^{\mu\nu}
	=	{\p^2\mc I\ov\p\Xi_\mu\p\Xi_\nu}
	&=	{\p^2 \abb{\Xi}\ov\p\Xi_\mu\p\Xi_\nu}{\df\mc I\ov\df\abb{\Xi}}
		+{\p \abb{\Xi}\ov\p\Xi_\mu}{\p \abb{\Xi}\ov\p\Xi_\nu}{\df^2\mc I\ov\df\abb{\Xi}^2}\nn
	&=	\paren{2\pi}^{d-1\ov2}\sqbr{
			\eta^{\mu\nu}{K_{d+1\ov2}\fn{\abb{\Xi}}\ov
				\abb{\Xi}^{{d+1\ov2}}}
			-{\Xi^\mu\Xi^\nu\ov \Xi^2}
				{K_{d+3\ov2}\fn{\abb{\Xi}}\ov \abb{\Xi}^{d-1\ov2}}
			}.
				\label{I mu nu}
}
By construction, we have an identity
\al{
\eta_{\mu\nu}\mc I^{\mu\nu}
	&=	-\mc I,
}
namely,
\al{
\pn{d+1}{K_{d+1\ov2}\fn{\abb\Xi}\ov\abb\Xi}
-K_{d+3\ov2}\fn{\abb\Xi}
	&=	-K_{d-1\ov2}\fn{\abb\Xi},
		\label{Bessel identity}
}
which is exactly the identity that the modified Bessel function satisfies.

\subsection{Integral of master integral}

For later convenience, we define
\al{
\mc I_{(n)}^{\nu_1\dots\nu_\ell}\fn{\Xi}
	&:=	\int{\df^d\bs u\ov2u^0}e^{u\cdot\Xi}{u^{\nu_1}\dots u^{\nu_\ell}\ov \pn{u^0}^n}
		\label{I integral defined}
}
for an arbitrary complex $D$-vector $\Xi$ that is not necessarily on-shell, while $u$ as always is: $u^0=\sqrt{1+\bs u^2}$. 
This integral is Lorentz-covariant when and only when $n=0$.
Note that $\mc I^{\nu_1\dots\nu_\ell}_{(0)}\fn{\Xi}=\mc I^{\nu_1\dots\nu_\ell}\fn{\Xi}$
and $\mc I_{(0)}\fn{\Xi}=\mc I\fn{\Xi}$.\footnote{
Comparing with the previous language,
$\Braket{\h p^{\mu_1}\cdots\h p^{\mu_n}}
	=	m^n\mc I^{\mu_1\dots\mu_n}\fn{2\sigma mP}/\mc I\fn{2\sigma mP}$.
}
We note that
\al{
\mc I_{(n)}^{\nu_1\dots\nu_\ell}\fn{\Xi}
	&=	{\p^\ell\ov\p\Xi_{\nu_1}\cdots\p\Xi_{\nu_\ell}}\mc I_{(n)}\fn{\Xi},
}
where indices are not summed.\footnote{
There also exists the relation ${\p^k\ov{\p\Xi_0}^k}\mc I_{(n)}^{\nu_1\dots\nu_\ell}\fn{\Xi}
	=	\mc I_{(n-k)}^{\nu_1\dots\nu_\ell}\fn{\Xi}$.
}

\subsubsection{Saddle-point method}
We compute $\mc I_{(n)}\fn{\Xi}$ in the large $\abb\Xi$ expansion. Concretely, we compute the following integral using the saddle-point method in the limit $\lambda\gg1$:
\al{
\mc I_{(n)}\fn{\lambda\Theta}
	&=	\int{\df^d\bs u\ov2u^0}e^{\lambda u\cdot\Theta}{1\ov\paren{u^0}^n}
	=	\int{\df^d\bs u\ov2}\, e^F,
}
where
\al{
F	&:=	\lambda u\cdot\Theta-\paren{n+1}\ln u^0.
}
Note that $\Theta$ is ``off-shell'' in the sense that $\Theta^0$ is treated as an independent variable.
For reference, we list the derivatives of exponent before setting to the saddle point:
\al{
{\p F\ov\p u^i}
	&=	\lambda\paren{\Theta_i-{u_i\ov u^0}\Theta^0}
		-\paren{n+1}{u_i\ov\paren{u^0}^2},\\
{\p^2F\ov\p u^i\p u^j}
	&=	-\delta_{ij}
			\paren{\lambda{\Theta^0\ov u^0}+{n+1\ov\pn{u^0}^2}}
		+u_iu_j
			\paren{\lambda{\Theta^0\ov\pn{u^0}^3}
				+2{n+1\ov\pn{u^0}^4}},
			\label{Hessian}\\
{\p^3F\ov\p u^i\p u^j\p u^k}
	&=	\pn{\delta_{ij}u_k+\delta_{jk}u_i+\delta_{ki}u_j}\pn{
			{\lambda\Theta^0\ov\pn{u^0}^3}
			+{2\pn{n+1}\ov\pn{u^0}^4}
			}
		-u_iu_ju_k\pn{
			{3\lambda\Theta^0\ov\pn{u^0}^5}
			+{8\pn{n+1}\ov\pn{u^0}^6}
			},\\
{\p^4F\ov\p u^i\p u^j\p u^k\p u^\ell}
	&=	\pn{\delta_{ij}\delta_{k\ell}+\delta_{jk}\delta_{i\ell}+\delta_{ki}\delta_{j\ell}}
			\pn{{\lambda\Theta^0\ov\pn{u^0}^3}
				+{2\pn{n+1}\ov\pn{u^0}^4}}\nn
	&\quad
		-\pn{
			\delta_{ij}u_ku_\ell+\delta_{jk}u_iu_\ell+\delta_{ki}u_ju_\ell
			+\delta_{i\ell}u_ju_k+\delta_{j\ell}u_iu_k+\delta_{k\ell}u_iu_j
			}
		\pn{
			{3\lambda\Theta^0\ov\pn{u^0}^5}
			+{8\pn{n+1}\ov\pn{u^0}^6}
			}\nn
	&\quad
		+u_iu_ju_ku_\ell\pn{
			{15\lambda\Theta^0\ov\pn{u^0}^7}
			+{48\pn{n+1}\ov\pn{u^0}^8}
			},
}
and the eigenvalues of the Hessian~\eqref{Hessian} are $\paren{d-1}$-fold degenerate
$-\lambda{\Theta^0\ov u^0}-{n+1\ov\paren{u^0}^2}$
and a single
$-{\lambda\Theta^0\ov\pn{u^0}^3}
		+\pn{n+1}
			\pn{
			{1\ov\pn{u^0}^2}-{2\ov\pn{u^0}^4}
			}$.

To obtain the saddle point, we solve
\al{
{\p F\ov\p u^i}
	&=	\lambda\paren{\Theta_i-{u_i\ov u^0}\Theta^0}
		-\paren{n+1}{u_i\ov\paren{u^0}^2}
	\sr{!}{=}
		0.\label{saddle point equation more precise}
}
We put an ansatz
\al{
u_*^i
	&=	\Theta^i\sum_{n=0}^\infty{C_n\ov\lambda^n}
}
to get the solution
\al{
C_0	&=	{1\ov\abb{\Theta}},&
C_1	&=	-{n+1\ov\abb{\Theta}^2},&
C_2	&
	=	{\paren{n+1}^2\ov2\abb{\Theta}^3}\paren{
			1
			+{\abb{\Theta}^2\ov\pn{\Theta^0}^2}
			},&
C_3
	&=	-{\paren{n+1}^3\ov\pn{\Theta^0}^4},
}
etc.\footnote{
Without large $\lambda$ expansion, we may directly put $u^i=C\Theta^i$:
\als{
\lambda\paren{1-C{\Theta^0\ov\sqrt{1+C^2\bs\Theta^2}}}-\paren{n+1}{C\ov1+C^2\bs\Theta^2}
	&=	0.
}
We may write down analytic solutions for $C$ explicitly, but we show the results for large $\lambda$ since the analytic one is too lengthy.
}
Then at the saddle point,
\al{
e^{F_*}
	&
	=	e^{-\lambda\abb{\Theta}}\paren{\abb{\Theta}\ov\Theta^0}^{n+1}
		\pn{
		1+{\pn{n+1}^2\ov2\lambda\abb{\Theta}}{\bs\Theta^2\ov\pn{\Theta^0}^2}
		+\cdots
		}.
}

Around the saddle point, we expand the integrand for large $\lambda$:
\al{
e^F
	&=	e^{F_*+{1\ov2}\Delta u_iM_{ij}\Delta u_j}
		\Bigg[
		1
		+{1\ov3!}{\p^3F_*\ov\p u^i\p u^j\p u^k}\Delta u_i\Delta u_j\Delta u_k
		+{1\ov4!}{\p^4F_*\ov\p u^i\p u^j\p u^k\p u^\ell}\Delta u_i\Delta u_j\Delta u_k\Delta u_\ell\nn
	&\phantom{=	e^{F_*+{1\ov2}\Delta u_iM_{ij}\Delta u_j}\Bigg(}
		+{1\ov2\,3!3!}\pn{{\p^3F_*\ov\p u^i\p u^j\p u^k}\Delta u_i\Delta u_j\Delta u_k}^2
		+\cdots
		\Bigg],
		\label{expanded exp F}
}
where
we have shifted the variables as $\Delta\bs u:=\bs u-\bs u_*$
and have neglected terms of order $1/\lambda^2$, with $\Delta\bs u$ being counted as $1/\sqrt\lambda$.
The derivatives at the saddle point are
\al{
\left.{\p^2F\ov\p u^i\p u^j}\right|_*
	&=	-\lambda\abb{\Theta}\paren{\delta_{ij}-{\Theta_i\Theta_j\ov\paren{\Theta^0}^2}}
		-\paren{n+1}\paren{\delta_{ij}-{\bs\Theta^2\ov\paren{\Theta^0}^2}{\Theta_i\Theta_j\ov\paren{\Theta^0}^2}}
		+\cdots\nn
	&=	-\lambda\abb{\Theta}\delta_{ij}\pn{
			1+{n+1\ov\lambda\abb{\Theta}}+\cdots
			}
		+\lambda\abb{\Theta}{\Theta_i\Theta_j\ov\pn{\Theta^0}^2}\pn{
			1+{n+1\ov\lambda\abb{\Theta}}{\bs\Theta^2\ov\pn{\Theta^0}^2}+\cdots
			},
		\label{Hessian expanded}
		\\
\left.{\p^3F\ov\p u^i\p u^j\p u^k}\right|_*
	&=	\pn{\delta_{ij}\Theta_k+\delta_{jk}\Theta_i+\delta_{ki}\Theta_j}
		\pn{\lambda{\abb\Theta^2\ov\pn{\Theta^0}^2}+\cdots}
		+\Theta_i\Theta_j\Theta_k\pn{-3\lambda{\abb\Theta^2\ov\pn{\Theta^0}^4}+\cdots},\\
\left.{\p^4F\ov\p u^i\p u^j\p u^k\p u^\ell}\right|_*
	&=	\pn{\delta_{ij}\delta_{k\ell}+\delta_{jk}\delta_{i\ell}+\delta_{ki}\delta_{j\ell}}
			\pn{
				{\lambda\abb\Theta^3\ov\pn{\Theta^0}^2}
				+\cdots}\nn
	&\quad
		-\pn{
			\delta_{ij}\Theta_k\Theta_\ell
			+\delta_{jk}\Theta_i\Theta_\ell
			+\delta_{ki}\Theta_j\Theta_\ell
			+\delta_{i\ell}\Theta_j\Theta_k
			+\delta_{j\ell}\Theta_i\Theta_k
			+\delta_{k\ell}\Theta_i\Theta_j
			}
			\pn{
			{3\lambda\abb\Theta^3\ov\pn{\Theta^0}^4}
			+\cdots
			}\nn
	&\quad
		+
		\Theta_i\Theta_j\Theta_k\Theta_\ell\pn{
			{15\lambda\abb\Theta^3\ov\pn{\Theta^0}^6}
			+\cdots
			}.
}

Let $M$ be the Hessian at the saddle point:
\al{
M_{ij}
	&:=	\left.{\p^2F\ov\p u^i\p u^j}\right|_*.
}
We write the eigenvectors and eigenvalues of $M$ as
\al{
MV^{(n)}
	&=	-\lambda^{(n)}V^{(n)},
}
in which $n$ is not summed.
Then we may diagonalize $M$ as
\al{
R^\t MR
	&=	\wt M
}
where
\al{
\wt M
	&:=	\bmat{-\lambda^{(1)}&&\\
			&\ddots&\\
			&&-\lambda^{(d)}
			},&
R	&:=	\bmat{V^{(1)}&\cdots&V^{(d)}}.
}
Here, $R$ is a complex orthogonal matrix $R^\t R=1$.

Now we may change the variables as
\al{
{1\ov2}\Delta u_iM_{ij}\Delta u_j
	&=	{1\ov2}\wt{\Delta u}_i\wt M_{ij}\wt{\Delta u}_j
	=	-{\lambda^{(n)}\ov2}\wt{\Delta u}_n^2,
}
with
\al{
\Delta u_i
	&=	R_{ij}\wt{\Delta u}_j.
	\label{change of variables}
}
Then we obtain
\al{
{1\ov3!}{\p^3F_*\ov\p u^i\p u^j\p u^k}\Delta u_i\Delta u_j\Delta u_k
	&=	{1\ov3!}\wt{\p^4F_*\ov\p u^i\p u^j\p u^k}\wt{\Delta u}_i\wt{\Delta u}_j\wt{\Delta u}_k,\\
{1\ov4!}{\p^4F_*\ov\p u^i\p u^j\p u^k\p u^\ell}\Delta u_i\Delta u_j\Delta u_k\Delta u_\ell
	&=	{1\ov4!}\wt{\p^4F_*\ov\p u^i\p u^j\p u^k\p u^\ell}\wt{\Delta u}_i\wt{\Delta u}_j\wt{\Delta u}_k\wt{\Delta u}_\ell,
}
where
\al{
\wt{\p^3F_*\ov\p u^i\p u^j\p u^k}
	&=	\pn{\delta_{ij}\wt\Theta_k+\delta_{jk}\wt\Theta_i+\delta_{ki}\wt\Theta_j}
		\pn{\lambda{\abb\Theta^2\ov\pn{\Theta^0}^2}+\cdots}
		+\wt\Theta_i\wt\Theta_j\wt\Theta_k
			\pn{-3\lambda{\abb\Theta^2\ov\pn{\Theta^0}^4}+\cdots},\\
\wt{\p^4F_*\ov\p u^i\p u^j\p u^k\p u^\ell}
	&=	\pn{\delta_{ij}\delta_{k\ell}+\delta_{jk}\delta_{i\ell}+\delta_{ki}\delta_{j\ell}}
			\pn{
				{\lambda\abb\Theta^3\ov\pn{\Theta^0}^2}
				+\cdots}\nn
	&\quad
		-\pn{
			\delta_{ij}\wt\Theta_k\wt\Theta_\ell
			+\delta_{jk}\wt\Theta_i\wt\Theta_\ell
			+\delta_{ki}\wt\Theta_j\wt\Theta_\ell
			+\delta_{i\ell}\wt\Theta_j\wt\Theta_k
			+\delta_{j\ell}\wt\Theta_i\wt\Theta_k
			+\delta_{k\ell}\wt\Theta_i\wt\Theta_j
			}
			\pn{
			{3\lambda\abb\Theta^3\ov\pn{\Theta^0}^4}
			+\cdots
			}\nn
	&\quad
		+
		\wt\Theta_i\wt\Theta_j\wt\Theta_k\wt\Theta_\ell\pn{
			{15\lambda\abb\Theta^3\ov\pn{\Theta^0}^6}
			+\cdots
			},
}

Because our Hessian takes the form
\al{
M_{ij}
	&=	A\delta_{ij}+B\Theta_i\Theta_j,
}
with
\al{
A	&=	-\lambda\abb\Theta\pn{1+{n+1\ov\lambda\abb\Theta}+\cdots},&
B	&=	\lambda{\abb\Theta\ov\pn{\Theta^0}^2}\pn{
			1+{n+1\ov\lambda\abb\Theta}{\bs\Theta^2\ov\pn{\Theta^0}^2}+\cdots
			},
}
we obtain
\al{
M_{kl}R_{ki}R_{lj}
	&=	A\delta_{ij}+B\wt\Theta_i\wt\Theta_j,
}
where $\wt\Theta_i:=\Theta_jR_{ji}$, that is, $\Theta_i=R_{ij}\wt\Theta_j$.
In particular,
\al{
\wt\Theta_k
	&=	\begin{cases}
		0		&	(k\neq d),\\
		\sqrt{\bs\Theta^2}	&	(k=d).
		\end{cases}
}
Then we get
\al{
\wt{\p^3F_*\ov\p u^i\p u^j\p u^k}\wt{\Delta u}_i\wt{\Delta u}_j\wt{\Delta u}_k
	&=	3\pn{\wt{\Delta\bs u}_\perp^2+\wt{\Delta u}_\parallel^2}
		\wt{\Delta u}_\parallel\sqrt{\bs\Theta^2}
		\pn{\lambda{\abb\Theta^2\ov\pn{\Theta^0}^2}+\cdots}\nn
	&\quad
		+\wt{\Delta u}_\parallel^3\pn{\bs\Theta^2}^{3\ov2}
			\pn{-3\lambda{\abb\Theta^2\ov\pn{\Theta^0}^4}+\cdots},\\
\wt{\p^4F_*\ov\p u^i\p u^j\p u^k\p u^\ell}\wt{\Delta u}_i\wt{\Delta u}_j\wt{\Delta u}_k\wt{\Delta u}_\ell
	&=	3\pn{\wt{\Delta\bs u}_\perp^2+\wt{\Delta u}_\parallel^2}^2
			\pn{
				{\lambda\abb\Theta^3\ov\pn{\Theta^0}^2}
				+\cdots}\nn
	&\quad
		-6\pn{\wt{\Delta\bs u}_\perp^2+\wt{\Delta u}_\parallel^2}\wt{\Delta u}_\parallel^2\bs\Theta^2
			\pn{
			{3\lambda\abb\Theta^3\ov\pn{\Theta^0}^4}
			+\cdots
			}\nn
	&\quad
		+
		\wt{\Delta u}_\parallel^4\pn{\bs\Theta^2}^2\pn{
			{15\lambda\abb\Theta^3\ov\pn{\Theta^0}^6}
			+\cdots
			},
}
where we have decomposed as $\wt{\Delta\bs u}=\wt{\Delta\bs u}_\perp+\wt{\Delta\bs u}_\parallel$ with
\al{
\wt{\Delta\bs u}_\perp
	&:=	\bmat{\wt{\Delta u}_1\\ \vdots\\ \wt{\Delta u}_{d-1}\\ 0},&
\wt{\Delta\bs u}_\parallel
	&:=	\bmat{0\\ \vdots\\ 0\\ \wt{\Delta u}_\parallel},
}
and used
\al{
\wt{\Delta\bs u}^2
	&=	\wt{\Delta\bs u}_\perp^2+\wt{\Delta u}_\parallel^2,&
\wt{\Delta\bs u}\cdot\wt{\bs\Theta}
	&=	\wt{\Delta u}_\parallel\sqrt{\bs\Theta^2}.
}

For later convenience, we define
\al{
I\fn{a,b}
	&:=	\int\df^d\wt{\Delta\bs u}\,
		e^{-{1\ov2}\lambda_l\wt{\Delta u_l}^2}
		\pn{\wt{\Delta\bs u}_\perp^2}^a
		\pn{\wt{\Delta u}_\parallel^2}^b.
}
The result is
\al{
I\fn{2,0}
	&=	\pn{2\pi\ov\lambda_\perp}^{d-1\ov2}
		\sqrt{2\pi\ov\lambda_\parallel}\,
		{d^2-1\ov\lambda_\perp^2},&
I\fn{2,1}
	&=	\pn{2\pi\ov\lambda_\perp}^{d-1\ov2}
		\sqrt{2\pi\ov\lambda_\parallel}\,
		{d^2-1\ov\lambda_\perp^2}
		{1\ov\lambda_\parallel},\\
I\fn{1,1}
	&=	\pn{2\pi\ov\lambda_\perp}^{d-1\ov2}
		\sqrt{2\pi\ov\lambda_\parallel}\,
		{d-1\ov\lambda_\perp\lambda_\parallel},&
I\fn{1,2}
	&=	\pn{2\pi\ov\lambda_\perp}^{d-1\ov2}
		\sqrt{2\pi\ov\lambda_\parallel}\,
		{d-1\ov\lambda_\perp}
		{3\ov\lambda_\parallel^2},\\
I\fn{0,2}
	&=	\pn{2\pi\ov\lambda_\perp}^{d-1\ov2}
		\sqrt{2\pi\ov\lambda_\parallel}\,
		{3\ov\lambda_\parallel^2},&
I\fn{0,3}
	&=	\pn{2\pi\ov\lambda_\perp}^{d-1\ov2}
		\sqrt{2\pi\ov\lambda_\parallel}\,
		{15\ov\lambda_\parallel^3},
}
where we have used, say,
\al{
I\fn{2,0}
	&=	\int\pn{\prod_{i=1}^{d-1}\df x_i}\df y\,e^{-{1\ov2}\pn{
		\lambda_\perp\sum_jx_j^2+\lambda_\parallel y^2}}
		\pn{\sum_{k=1}^{d-1}x_k^4+\sum_{k\neq\ell}x_k^2x_\ell^2},
}
as well as the one-dimensional integrals: $\int\df x\, e^{-{\lambda\ov2}x^2}x^2={\sqrt{2\pi}\ov\lambda^{3/2}}$, $\int\df x\, e^{-{\lambda\ov2}x^2}x^4={3\sqrt{2\pi}\ov\lambda^{5/2}}$, and $\int\df x\, e^{-{\lambda\ov2}x^2}x^6={15\sqrt{2\pi}\ov\lambda^{7/2}}$.

In the expansion~\eqref{expanded exp F}, the cubic and quintic integrals vanish, while the quartic and hexic ones become
\al{
&\int\df^d\wt{\Delta\bs u}\,e^{-{1\ov2}\lambda_l\wt{\Delta u_l}^2}\wt{\p^4F_*\ov\p u^i\p u^j\p u^k\p u^\ell}\wt{\Delta u}_i\wt{\Delta u}_j\wt{\Delta u}_k\wt{\Delta u}_\ell\nn
	&\quad=	{3\lambda\abb\Theta^3\ov\pn{\Theta^0}^2}
			\pn{
				I\fn{2,0}+2I\fn{1,1}+I\fn{0,2}
				-{6\bs\Theta^2\ov\pn{\Theta^0}^2}
					\Pn{I\fn{1,1}+I\fn{0,2}}
				+{5\pn{\bs\Theta^2}^2\ov\pn{\Theta^0}^4}
					I\fn{0,2}
				}
			+\cdots\nn
	&\quad=
		\pn{2\pi\ov\lambda_\perp}^{d-1\ov2}
		\sqrt{2\pi\ov\lambda_\parallel}
		{3\lambda\abb\Theta^3\ov\pn{\Theta^0}^2}
		\Bigg[
			\pn{
				{d^2-1\ov\lambda_\perp^2}
				+{2\pn{d-1}\ov\lambda_\perp\lambda_\parallel}
				+{3\ov\lambda_\parallel^2}
				}
			-{6\bs\Theta^2\ov\pn{\Theta^0}^2}\pn{
				{d-1\ov\lambda_\perp\lambda_\parallel}
				+{3\ov\lambda_\parallel^2}
				}
			+{\pn{\bs\Theta^2}^2\ov\pn{\Theta^0}^4}
			{15\ov\lambda_\parallel^2}
			+\cdots
		\Bigg],
		\label{4th order term}\\
&\int\df^d\wt{\Delta\bs u}\,e^{-{1\ov2}\lambda_l\wt{\Delta u_l}^2}
	\pn{\wt{\p^3F_*\ov\p u^i\p u^j\p u^k}\wt{\Delta u}_i\wt{\Delta u}_j\wt{\Delta u}_k}^2\nn
	&\quad=	9\bs\Theta^2\lambda^2{\abb\Theta^4\ov\pn{\Theta^0}^4}\Bigg(
			I\fn{2,1}
			+2{\abb\Theta^2\ov\pn{\Theta^0}^2}I\fn{1,2}
			+{\abb\Theta^4\ov\pn{\Theta^0}^4}I\fn{0,3}
			\Bigg)
		+\cdots\nn
	&\quad=	9\bs\Theta^2\lambda^2\pn{2\pi\ov\lambda_\perp}^{d-1\ov2}
		\sqrt{2\pi\ov\lambda_\parallel}
		{\abb\Theta^4\ov\pn{\Theta^0}^4}\Bigg(
			{d^2-1\ov\lambda_\perp^2\lambda_\parallel}
			+2{\abb\Theta^2\ov\pn{\Theta^0}^2}
				{3\pn{d-1}\ov\lambda_\perp\lambda_\parallel^2}
			+{\abb\Theta^4\ov\pn{\Theta^0}^4}
				{15\ov\lambda_\parallel^3}
			\Bigg)
		+\cdots.
}
As said above, $\wt{\p^6F_*\ov\p u^i\p u^j\p u^k\p u^l\p u^m\p u^n}\wt{\Delta u}_i\wt{\Delta u}_j\wt{\Delta u}_k\wt{\Delta u}_l\wt{\Delta u}_m\wt{\Delta u}_n$ is $\Or{\lambda^{-2}}$ and does not contribute to the order of our interest.

Concretely, the eigenvalues of $M$ are $\pn{d-1}$-fold degenerate $-\lambda_\perp$ and a single $-\lambda_\parallel$, where
\al{
\lambda_\perp
	&:=	\lambda\abb{\Theta}\sqbr{
		1+{n+1\ov\lambda\abb{\Theta}}
		+\cdots
		}&
&\tx{for $\lambda^{(n)}$ with $n=1,\dots,d-1$},\\
\lambda_\parallel
	&:=	\lambda{\abb{\Theta}^3\ov\paren{\Theta^0}^2}
		\sqbr{
		1
		+{n+1\ov\lambda\abb{\Theta}}\pn{1+{\bs\Theta^2\ov\pn{\Theta^0}^2}}
		+\cdots
		}&
&\tx{for $\lambda^{(n)}$ with $n=d$}.
}
Keeping the lowest order terms for large $\lambda$ in Eq.~\eqref{4th order term}, we get
\al{
\int\df^d\wt{\Delta\bs u}\,e^{-{1\ov2}\lambda_l\wt{\Delta u_l}^2}\wt{\p^4F_*\ov\p u^i\p u^j\p u^k\p u^\ell}\wt{\Delta u}_i\wt{\Delta u}_j\wt{\Delta u}_k\wt{\Delta u}_\ell
	&=	\pn{2\pi\ov\lambda\abb\Theta}^{d\ov2}{3\Theta^0\ov\lambda}
		\pn{d+2}\pn{{d+4\ov\pn{\Theta^0}^2}-{4\ov\abb\Theta^2}}
		+\cdots,\\
\int\df^d\wt{\Delta\bs u}\,e^{-{1\ov2}\lambda_l\wt{\Delta u_l}^2}
	\pn{\wt{\p^3F_*\ov\p u^i\p u^j\p u^k}\wt{\Delta u}_i\wt{\Delta u}_j\wt{\Delta u}_k}^2
	&=	\pn{2\pi\ov\lambda\abb\Theta}^{d\ov2}9(d+2)(d+4)
			{\bs\Theta^2\ov\abb\Theta^2}{1\ov\lambda\Theta^0}
		+\cdots.
}

To summarize,
\al{
\mc I_{(n)}
	&=	{1\ov2}\paren{2\pi\ov\lambda\abb{\Theta}}^{d\ov2}
		e^{-\lambda\abb{\Theta}}
		\pn{\abb{\Theta}\ov\Theta^0}^n\nn
	&\quad\times
		\sqbr{
			1+{1\ov\lambda\abb\Theta}\pn{
				{d\pn{d-2}\ov8}
				+{n\ov2}\pn{-d+\pn{n+1}{\bs\Theta^2\ov\pn{\Theta^0}^2}}
				}
			+\cdots
			},
}
and hence
\al{
{1\ov\mc I_{(0)}}{\p\mc I_{(n)}\ov\lambda\p\Theta^i}
	&=	{\Theta_i\ov\abb\Theta}\pn{\abb\Theta\ov\Theta^0}^n\sqbr{
			1+{1\ov2\lambda\abb\Theta}\pn{
					d
					-n\pn{d+2}+n\pn{n+1}{\bs\Theta^2\ov\pn{\Theta^0}^2}
					}
			+\Or{1\ov\lambda^2}
			},
				\label{saddle-point 1st}\\
{1\ov\mc I_{(0)}}{\p^2\mc I_{(n)}\ov\lambda^2\p\Theta^i\p\Theta^j}
	&=	\pn{\abb\Theta\ov\Theta^0}^n
		\Bigg[
			{\Theta_i\Theta_j\ov\abb\Theta^2}
			+{1\ov\lambda\abb\Theta}\nn
	&\qquad\times
		\br{
				\delta_{ij}
				+{\Theta_i\Theta_j\ov\abb\Theta^2}\sqbr{
					d+1
					-{n\ov2}\pn{d+4-\pn{n+1}{\bs\Theta^2\ov\pn{\Theta^0}^2}}
					}
				}
			+\Or{1\ov\lambda^2}
		\Bigg],
		\label{saddle-point 2nd}
}
where we used
\al{
{\p\abb\Theta\ov\p\Theta^i}
	&=	{\p\sqrt{\pn{\Theta^0}^2-\bs\Theta^2}\ov\p\Theta^i}
	=	-{\Theta^i\ov\abb\Theta}.
}

\subsubsection{Non-relativistic expansion without using saddle-point method}
Instead of using the saddle-point method, we may obtain the integral~\eqref{I integral defined} by the non-relativistic expansion, which provides a non-trivial consistency check.

In the non-relativistic limit $\Re\Xi^0\gg\ab{\Re\bs\Xi}\geq0$, the dominant contribution to the integral would be from the non-relativistic velocity $\ab{\bs u}\ll 1$:
\al{
\mc I_{(n)}\fn{\Xi}
	&=	\int{\df^d\bs u\ov2u^0}e^{u\cdot\Xi}{1\ov\paren{u^0}^n}\nn
	&=	\int{\df^d\bs u\ov2u^0}e^{u\cdot\Xi}
			\paren{1-{n\ov2}\bs u^2+{n\pn{n+2}\ov8}\pn{\bs u^2}^2+\cdots}\nn
	&=	\mc I\fn{\abb\Xi}
		-{n\ov2}{\p^2\ov\p\Xi^i\p\Xi^i}\mc I\fn{\abb\Xi}
		+{n\pn{n+2}\ov8}{\p^4\ov\p\Xi^i\p\Xi^i\p\Xi^j\p\Xi^j}\mc I\fn{\abb\Xi}
		+\cdots\nn
	&=	\pn{2\pi}^{d-1\ov2}{K_{d-1\ov2}\fn{\abb\Xi}\ov\abb{\Xi}^{d-1\ov2}}\sqbr{
			1
			-{n\ov2}\pn{
				{d\,K_{d+1\ov2}\fn{\abb\Xi}\ov\abb\Xi K_{d-1\ov2}\fn{\abb\Xi}}
				+{\bs\Xi^2K_{d+3\ov2}\fn{\abb\Xi}\ov\abb{\Xi}^2K_{d-1\ov2}\fn{\abb\Xi}}
				}
			+\cdots
			}.
}
where we used
\al{
{\p\abb\Xi\ov\p\Xi^i}
	&=	-{\Xi_i\ov\abb\Xi},&
{\p^2\abb\Xi\ov\p\Xi^i\p\Xi^j}
	&=	-{\delta_{ij}\ov\abb\Xi}
		-{\Xi_i\Xi_j\ov\abb\Xi^3}.
}
In the above, $\Xi^0$ are written in terms of $\abb\Xi$ and $\bs\Xi$.

Now we show results in the non-relativistic expansion with large $\abb\Xi$, leaving the overall function that cancels out in ${1\ov\mc I_{(0)}}{\p\mc I_{(n)}\ov\p\Xi^i}$, etc.:
\al{
\mc I_{(n)}
	&=	\pn{2\pi}^{d-1\ov2}{K_{d-1\ov2}\fn{\abb\Xi}\ov\abb{\Xi}^{d-1\ov2}}
			\pn{
				1-{dn\ov2\abb{\Xi}}+\cdots
				},\\
{\p\mc I_{(n)}\ov\p\Xi^i}
	&=	\pn{2\pi}^{d-1\ov2}{K_{d-1\ov2}\fn{\abb\Xi}\ov\abb{\Xi}^{d-1\ov2}}
			{\Xi_i\ov\abb\Xi}
			\pn{
				1+{d-n\pn{d+2}\ov2\abb\Xi}+\cdots
				},\\
{\p^2\mc I_{(n)}\ov\p\Xi^i\p\Xi^j}
	&=	\pn{2\pi}^{d-1\ov2}{K_{d-1\ov2}\fn{\abb\Xi}\ov\abb{\Xi}^{d-1\ov2}}
		{1\ov\abb\Xi}\sqbr{
			\delta_{ij}
			+{1\ov\abb\Xi}\pn{
				{d-n\pn{d+2}\ov2}\delta_{ij}+\Xi_i\Xi_j
				}
			+\cdots
			}.
}

As a cross check, we put $\Xi=\lambda\Theta$ and take the saddle-point limit for large $\lambda$:
\al{
{1\ov\mc I_{(0)}}{\p\mc I_{(n)}\ov\lambda\p\Theta^i}
	&=	{\Theta_i\ov\abb\Theta}\sqbr{
			1+{d-n\pn{d+2}\ov2\lambda\abb\Theta}+\Or{1\ov\lambda^2}
			},\\
{1\ov\mc I_{(0)}}{\p^2\mc I_{(n)}\ov\lambda^2\p\Theta^i\p\Theta^j}
	&=	{\Theta_i\Theta_j\ov\abb\Theta^2}
		+{1\ov\lambda\abb\Theta}\pn{
			\delta_{ij}
			+{\Theta_i\Theta_j\ov\abb\Theta^2}
				\pn{d+1-{n\ov2}\pn{d+4}}
			}
		+\Or{1\ov\lambda^2}.
}
We can confirm that these results coincide with the saddle-point ones~\eqref{saddle-point 1st} and \eqref{saddle-point 2nd} if we take the NR limit ${\abb\Theta\ov\Theta^0}\to1$ and ${\bs\Theta^2\ov\pn{\Theta^0}^2}\to0$ there.


\bibliographystyle{JHEP}
\bibliography{refs}

\end{document}